\documentclass{article}
\pdfoutput=1
\usepackage{PRIMEarxiv}
\usepackage[utf8]{inputenc} 
\usepackage[T1]{fontenc}    
\usepackage{hyperref}       
\usepackage{url}            
\usepackage{booktabs}       
\usepackage{nicefrac}       
\usepackage{microtype}      
\usepackage{lipsum}
\usepackage{fancyhdr}       
\usepackage[pdftex]{graphicx} 
\usepackage{epstopdf}
\usepackage{amsmath,amsfonts,amssymb,xcolor,mathtools}
\usepackage{mathrsfs}
\newcommand{\upi}{\mathrm{i}}
\newcommand{\upd}{\mathrm{d}}
\newcommand{\intinf}{\int_{-\infty}^\infty}
\newcommand{\intL}{\int_{-L}^\infty}
\newcommand{\sgn}{\mathrm{sgn}}
\renewcommand{\Re}{\mathrm{Re}}
\DeclareMathAlphabet{\mathsfbi}{OT1}{\sfdefault}{bx}{sl}
  
\title{Generalised eigenfunction expansion and singularity expansion methods for canonical time-domain wave scattering problems
}

\author{
  Ben Wilks\\
  Department of Mathematics and Statistics\\
  University of Otago\\
  Dunedin\\
  New Zealand\\
  \texttt{wilbe612@student.otago.ac.nz}\\
  \AND
  Michael H. Meylan \\
  School of Information and Physical Sciences\\
  The University of Newcastle\\
  Newcastle\\
  Australia\\
  \AND
  Fabien Montiel\\
  Department of Mathematics and Statistics\\
  University of Otago\\
  Dunedin\\
  New Zealand\\
  \AND
  Sarah Wakes\\
  Department of Mathematics and Statistics\\
  University of Otago\\
  Dunedin\\
  New Zealand\\
}

\newcommand{\red}[1]{\textcolor{black}{#1}}

\usepackage{soul}

\begin{document}
\maketitle

\begin{abstract}
The generalised eigenfunction expansion method (GEM) and the singularity expansion method (SEM) are applied to solve the canonical problem of wave scattering on an infinite stretched string in the time domain. The GEM, which is shown to be equivalent to d'Alembert's formula when no scatterer is present, is also derived in the case of a point-mass scatterer coupled to a spring. The discrete GEM, which generalises the discrete Fourier transform, is shown to reduce to matrix multiplication. The SEM, which is derived from the Fourier transform and the residue theorem, is also applied to solve the problem of scattering by the mass-spring system. The GEM and SEM are also used to solve the problem of wave scattering by a mass positioned a fixed distance from an anchor point, which supports more complicated resonant behavior. 
\end{abstract}

\section{Introduction}
The majority of the contemporary wave scattering literature is concerned with solving the frequency-domain problem, i.e., finding time-harmonic solutions of the associated wave equation. Frequency domain problems are very well understood and there are numerous methods for solving them \cite{martin2006}. However, the corresponding time domain problems are addressed less frequently. This is unfortunate, because solving the time-domain problem and visualising the output often provides additional insights which cannot be anticipated from the frequency-domain solutions alone. For example, it can be difficult to see features such as mode shapes, resonances, and decay patterns. Moreover, the frequency-domain solutions on unbounded domains are unrealistic because they require infinite time and infinite energy to excite. That said, the frequency-domain solutions are important because they encode the diagonalising transformation \red{(i.e., the spectral decomposition of the problem)} to solve the time-dependent problem. This is revealed by the generalised eigenfunction expansion method (GEM).

\red{Let us first recall} the advantages of using diagonalising transformations in contrast to integral transforms when solving linear systems of ordinary differential equations (ODE). Consider the initial value problem
\begin{subequations}\label{matrix_ODE}
  \begin{align}
    \frac{\upd\mathbf{x}}{\upd t}&=A\mathbf{x},\\
    \mathbf{x}(0) &= \mathbf{x}_0,
\end{align}  
\end{subequations}
where $\mathbf{x}(t)\in\mathbb{R}^n$ for all $t\geq 0$, $A\in M_{n\times n}(\mathbb{R})$ is diagonalisable and $\mathbf{x}_0\in\mathbb{R}^n$. Let $\lambda_j$ and $\mathbf{v}_j$ denote the eigenvalues and eigenvectors of $A$ and assume that $\{\mathbf{v}_j|1\leq j\leq n\}$ is an orthonormal basis of $\mathbb{R}^n$. Then the solution of \eqref{matrix_ODE} can \red{be solved using a spectral decomposition of $A$, yielding}
\begin{equation}\label{matrix_ODE_spectral_sol}
    \mathbf{x}(t)=\sum_{j=1}^n \langle \mathbf{x}_0,\mathbf{v}_j\rangle e^{\lambda_j t}\mathbf{v}_j,
\end{equation}
\red{where $\langle\cdot,\cdot\rangle$ is the usual dot product.} Equation \eqref{matrix_ODE} can also be solved using \red{the} Laplace transform, which yields the more cumbersome formula
\begin{equation}
    \mathbf{x}(t)=\mathcal{L}^{-1}\left\{(sI-A)^{-1}\mathbf{x}_0\right\},
\end{equation}
where $\mathcal{L}^{-1}$ denotes the inverse Laplace transform, $s$ denotes the Laplace transform variable and $I$ is the $n\times n$ identity matrix. \red{Equation \eqref{matrix_ODE} can also be solved using an iterative method, such as Euler's method or a Runge-Kutta method. The main disadvantage of such schemes is that the error grows in time. This does not occur in the spectral approach, where the error only depends on the accuracy of the diagonalisation and the numerical error of evaluating \eqref{matrix_ODE_spectral_sol}, not on the time at which the solution is sought.}

\red{The GEM extends the spectral decomposition \eqref{matrix_ODE_spectral_sol} to scalar linear wave scattering problems, a wide class of which can be written in the form
\begin{subequations}\label{first_order_system_intro}    
\begin{align}
    \upi \partial_t\mathbf{u}(x,t)&=\mathscr{P}\mathbf{u}(x,t),\\
    \mathbf{u}(x,0)&=\mathbf{f}(x),
\end{align}
\end{subequations}
where $\mathscr{P}$ is a differential operator describing the time evolution of the system, and $\mathbf{u}$ belongs to a space of two-component functions which satisfy the boundary conditions. Here, $x$ is the spatial variable, which we restrict to a subset of $\mathbb{R}$ in this paper (i.e.\ one spatial dimension). Some examples showing how canonical wave scattering problems can be rewritten in this form are given in 
\textsection\textsection \ref{mass-spring_sec} and \ref{Anchor_point_sec}.}

\red{Before discussing the advantages of the GEM, we consider the solution of \eqref{first_order_system_intro} using more familiar methods, i.e.\ the Fourier transform method and time-stepping approaches. We define the Fourier transform and its inverse as
\begin{equation}
    \widehat{h}(\omega) = \int_0^\infty h(t)e^{\upi\omega t}\upd t\quad\text{and}\quad h(t)=\frac{1}{2\pi}\intinf \widehat{h}(\omega)e^{-\upi\omega t}\upd \omega,
\end{equation}
respectively, where $h(t)$ is assumed to vanish for $t<0$. This is equivalent to the Laplace transform with the change of variables $s=-\upi\omega$. After taking the Fourier transform of \eqref{first_order_system_intro}, we obtain
\begin{equation}\label{inhomogeneous_helmholtz_intro}
    \upi (\mathscr{P}-\omega)\widehat{\mathbf{u}}(x,\omega)=\mathbf{f}(x),
\end{equation}
which is equivalent to an inhomogeneous Helmholtz equation. The GEM is more straightforward to apply because it only requires the solution to the corresponding homogeneous problem.}

\red{An iterative method (such as the finite-difference-time-domain method) could also be used to solve \eqref{first_order_system_intro}. However, as is the case for iterative solutions of ODE \eqref{matrix_ODE}, this introduces an error which grows with time. The GEM provides a practical solution to the drawbacks of the two approaches discussed here.}

The GEM, which was introduced by Povzner \cite{povzner1953expansion} and further developed by Ikebe \cite{ikebe1960eigenfunction} for the Schr\"{o}dinger equation, is a powerful technique for solving time-domain wave scattering problems. It has previously been applied to two- and three-dimensional water wave scattering  \red{\cite{hazard2007generalized,hazard2008spectral,meylan2009time,meylan_2009,peter2010general,meylan_fitzgerald_2014}}. The GEM is perhaps the most direct path from the frequency domain to the time domain, \red{because it involves a spectral decomposition of the time evolution operator, analogous to \eqref{matrix_ODE_spectral_sol}, whose eigenfunctions are precisely the frequency-domain scattering solutions. The GEM can only be applied when the time evolution operator is self-adjoint, which rules out systems with dissipation.} The GEM is straightforward to implement numerically, as the solution reduces to a linear integral expression, which can then be turned into matrix multiplication after applying a quadrature rule \red{(this observation has not been published before, although see \cite{Meylan2023MWSW03})}. It is therefore surprising that time-domain solutions are rarely explored by researchers interested in wave scattering problems, especially as they typically already have the frequency-domain solutions required to implement the GEM. This suggests that the techniques behind the GEM are not widely known or understood. The treatment of the GEM in the existing literature is at least partially responsible for its lack of adoption--it is often applied to complex, field-specific problems and, to our knowledge, has never been explained in a simple setting. Furthermore, it was not discussed in an otherwise comprehensive recent textbook on time-domain scattering \cite{martin2021time}. In this paper, we seek to remedy this by illustrating the method in the context of a simple canonical wave-scattering problem.

\red{We also consider the singularity expansion method (SEM) in the canonical context of this paper.} This method, which was developed by Baum \cite{baum1971singularity,baum2005singularity} after observing that the transient electromagnetic scattering of a pulse by an aircraft is dominated by a collection of damped sinusoids, has found applications in target identification \cite{Baum_1997}. The method seeks to approximate the time-domain solution as an expansion over the unforced modes of the problem. These modes, which are also known as complex resonances, are homogeneous scattering solutions to problems in unbounded domains (i.e., nontrivial solutions with no incoming wave) that generally occur at complex frequencies \cite{pagneux2013trapped,Kristensen2020}. Trapped modes (i.e. unforced modes occurring at real frequencies) are special cases of complex resonances which have finite energy, whereas complex resonances at non-real frequencies are unbounded in the far field. By expanding the solution as a superposition of the complex resonance solutions, the SEM extends the notion of expressing the transient behaviour of a bounded system as an expansion over its natural modes\red{,} to unbounded domains. In systems with long-lived resonant modes, the SEM becomes accurate at large times but is inaccurate at small times, except in special cases, i.e. those that satisfy the requirements of Lax-Phillips scattering theory \cite{meylan_2002}, the formal mathematical \red{framework} for studying scattering operators. \red{The convergence properties of the SEM are not well understood in general, as it remains unclear how the infinite modal expansions should be truncated in systems with infinitely many complex resonances (although see \cite{baldassari2021modal}). Moreover, the normalisation of the complex resonances is a major challenge for implementing the SEM \cite{meylan_fitzgerald_2014}. We show that this normalisation is elementary in the problems we consider here.}

In this paper, we consider the scattering of transverse waves on a stretched string by a point scatterer, which in this paper will be either a mass or a mass-spring system. The physics of this problem is governed by the one-dimensional wave equation with a matching condition at the position of the scatterer \cite{billingham2000wave}. With suitable adjustments, this also models acoustic waves in thin tubes via the single-mode approximation, and water waves in shallow water. Due to its simplicity, the one-dimensional wave equation is the ideal starting point for disseminating the GEM and SEM. Other recent studies of one-dimensional scattering have improved the understanding of complex wave scattering phenomena, which can then inform more realistic physical models, including studies of wave scattering by periodic media \cite{MARTIN20142_N_masses,MARTIN2015_semi-infinite}, graded metamaterial devices \cite{porter2018waves,davies2023problem} and ice shelves \cite{bennetts2021complex}.

The outline of this paper is as follows. In \textsection\ref{dAlembert_sec} we show that the GEM can be used to derive the well-known d'Alembert solution of the wave equation on an infinite string. In \textsection\ref{mass-spring_sec} we augment the string with a resonant mass-spring point scatterer. We find the time-domain solution of this problem using the GEM by expanding it as a continuous superposition of frequency-domain problems with left-travelling and right-travelling incident waves. Subsequently, we approximate the solution numerically using midpoint quadrature, which allows us to write a simple expression for the GEM solution in terms of matrix multiplication. In \textsection\ref{SEM_sec}, we derive the SEM formula using the Fourier transform and the residue theorem. We apply the SEM to the problem of wave scattering by a mass-spring system introduced in \textsection\ref{mass-spring_sec}. To illustrate the GEM and SEM in a system with more complicated resonant behaviour, in \textsection\ref{Anchor_point_sec} we apply them to the \red{problem of scattering by a mass} positioned at a fixed distance from a point where the string is anchored. \red{In \textsection\ref{validation_sec}, we validate our results against a standard solution method for one-dimensional scattering problems, namely, one that uses an ansatz consisting of incoming and outgoing waves to reduce the problem to an ODE. We give a} brief conclusion in \textsection\ref{conclusion_sec}. In order to demonstrate how readily the GEM and SEM can be implemented, \red{the reader is encouraged to view the} fully documented MATLAB code for the problems considered in this paper\red{, which} are provided in the supplementary material. The code is provided both as MATLAB live scripts and as PDF files. The supplementary material also contains animations corresponding \red{to all of the results presented} in this paper.

\section{Derivation of d'Alembert's formula using the generalised eigenfunction expansion}\label{dAlembert_sec}
\red{We begin by applying the GEM to solve the problem of wave propagation on an infinitely-long stretched string, eventually obtaining d'Alembert's formula. Although this may appear as a convoluted way of deriving a simple, classical result, this exercise will help us introduce key concepts and aspects of the GEM that will later be extended to scattering problems in which d'Alembert's formula does not apply.}

The string is assumed to have constant tension $T$ and constant mass density $\mu$. The displacement of the string $u$ is governed by the one-dimensional wave equation \cite{billingham2000wave}
\begin{subequations}\label{wave_equation_IVP}
    \begin{align}
    \partial_t^2 u-c^2\partial_x^2 u&=0\label{wave_equation}\\
    u(x,0)&=f(x)\label{ic1}\\
    \partial_t u(x,0)&=g(x),\label{ic2}
\end{align}
\end{subequations}
where $c=\sqrt{T/\mu}$. We assume that $f,\,g\to 0$ as $|x|\to\infty$. We rewrite \eqref{wave_equation} as a system of first-order equations in time
\begin{equation}\label{first_order_system}
    \upi \partial_t\begin{bmatrix}u\\ \upi v\end{bmatrix}=\mathscr{P}\begin{bmatrix}u\\ \upi v\end{bmatrix},
\end{equation}
where
\begin{equation}\label{L_def}
    \mathscr{P}=\begin{bmatrix}
        0&1\\ -c^2\partial_x^2&0
    \end{bmatrix}.
\end{equation}
Note that \eqref{first_order_system} has the form of a Schr\"{o}dinger equation, in which $\mathscr{P}$ is analogous to the Hamiltonian operator. To apply the GEM, we must choose an inner product that is conserved in time so that the operator $\mathscr{P}$ is self-adjoint. Namely, we require that the inner product satisfies
\begin{equation}
    \frac{\upd}{\upd t}\left\langle\begin{bmatrix}
        u\\ \upi v
    \end{bmatrix}(\cdot,t),\begin{bmatrix}
        u\\ \upi v
    \end{bmatrix}(\cdot,t)\right\rangle=0,
\end{equation}
for all $t$, where $[u,\upi v]^\intercal$ solves \eqref{first_order_system}. Conservation of energy suggests that inner products based on the total energy of the system are suitable candidates for this inner product. The total energy of the string is the sum of potential and kinetic energy, i.e.
\begin{equation}\label{string-energy}
    E(t)=\tfrac{1}{2}\intinf T|\partial_x u(x,t)|^2+\mu|v(x,t)|^2 \upd x.
\end{equation}
It is straightforward to show that the solutions of \eqref{first_order_system} conserve energy. Thus, we introduce an energy inner product of the form
\begin{equation}
    \left\langle\begin{bmatrix}
        u_1\\ \upi v_1
    \end{bmatrix},\begin{bmatrix}
        u_2\\ \upi v_2
    \end{bmatrix}\right\rangle \coloneqq \intinf T(\partial_x u_1)(\partial_x u_2)^* + \mu v_1v_2^* \upd x\red{,}
\end{equation}
which satisfies
\begin{equation}
    E = \frac{1}{2}\left\langle\begin{bmatrix}
        u\\ \upi v
    \end{bmatrix}(\cdot,t),\begin{bmatrix}
        u\\ \upi v
    \end{bmatrix}(\cdot,t)\right\rangle
\end{equation}
for all $t$. \red{The associated function space is}
\red{\begin{equation}\label{fun_space_1}
    \mathscr{H}=\left\{\left.\begin{bmatrix}
        u\\ \upi v
    \end{bmatrix}:\mathbb{R}\to\mathbb{C}^2\right|\left\langle\begin{bmatrix}
        u\\ \upi v
    \end{bmatrix},\begin{bmatrix}
        u\\ \upi v
    \end{bmatrix}\right\rangle<\infty\right\}.
\end{equation}}
To show that the operator $\mathscr{P}$ is, in fact, self-adjoint with respect to the energy inner product, integration by parts gives
\begin{align}
    \left\langle\mathscr{P}\begin{bmatrix}
        u_1\\ \upi v_1
    \end{bmatrix},\begin{bmatrix}
        u_2\\ \upi v_2
    \end{bmatrix}\right\rangle&=\intinf T(\partial_x \upi v_1)(\partial_x u_2)^*+\mu(\upi c^2\partial_x^2 u_1)(v_2)^*\upd x\nonumber\\
    &=\intinf -\upi T v_1 (\partial_x^2 u_2)^*-\upi T\partial_x u_1(\partial_x v_2)^*\upd x\nonumber\\
    &=\intinf \mu (\upi v_1) (-c^2\partial_x^2 u_2)^*+T\partial_x u_1(\partial_x \upi v_2)^*\upd x\nonumber\\
    &=\left\langle\begin{bmatrix}
        u_1\\ \upi v_1
    \end{bmatrix},\mathscr{P}\begin{bmatrix}
        u_2\\ \upi v_2
    \end{bmatrix}\right\rangle.\label{self-adjoint-proof}
\end{align}
As a consequence, $\mathscr{P}$ has real eigenvalues $\omega$ i.e.
\begin{equation}\label{fd}
    \mathscr{P}\begin{bmatrix}
        \breve{u}\\ \upi \breve{v}
    \end{bmatrix}=\begin{bmatrix}
        0&1\\ -c^2\partial_x^2&0
    \end{bmatrix}\begin{bmatrix}
        \breve{u}\\ \upi \breve{v}\end{bmatrix}=\omega\begin{bmatrix}
        \breve{u}\\ \upi \breve{v}\end{bmatrix}\red{,}
\end{equation}
\red{for some vector $[\breve{u},\upi \breve{v}]^{\mathrm{T}}\in\mathscr{H}$. Equation \eqref{fd}} is simply a restatement of the associated frequency-domain problem (with $\exp(-\upi \omega t)$ time-harmonic dependence), which is given by
\begin{subequations}
    \begin{align}
    \breve{v} &= -\upi \omega \breve{u}\\
    \partial_x^2\breve{u}&=-k^2\breve{u},\label{helmholtz}
\end{align}
\end{subequations}
where $k=\omega/c$. \red{In particular,} \eqref{helmholtz} is the one-dimensional Helmholtz equation whose general solution is a superposition of the left- and right-travelling frequency-domain solutions, which we write as
\begin{equation}\label{fd_sol_no_scatterer}
    \begin{bmatrix}
        \breve{u}_\pm\\\upi\breve{v}_\pm
    \end{bmatrix}
    =e^{\pm\upi kx}\begin{bmatrix}
        1\\ \upi(-\upi\omega)
    \end{bmatrix}.
\end{equation}
Note that these are generalised eigenfunctions of $\mathscr{P}$ because they do not have finite energy. Let us verify the orthogonality of the frequency-domain solutions with respect to the inner product. First, we show the orthogonality between the \red{left- and right-travelling} solutions at respective frequencies $\omega_j$ and wavenumbers $k_{j}=\omega_{j}/c$. We have
\begin{align}
    \left\langle\begin{bmatrix}
        \breve{u}_+\\ \upi \breve{v}_+
    \end{bmatrix}(\cdot,\omega_1),\begin{bmatrix}
        \breve{u}_-\\ \upi \breve{v}_-
    \end{bmatrix}(\cdot,\omega_2)\right\rangle&=\intinf\mu(-\upi\omega_1 e^{\upi k_1 x})(-\upi \omega_2e^{-\upi k_2x})^*\nonumber\\
    &\qquad+T(\upi k_1 e^{\upi k_1 x})(-\upi k_2 e^{-\upi k_2 x})^*\upd x\nonumber\\
    &=\intinf \left(\mu\omega_1\omega_2-Tk_1k_2\right)e^{\upi(k_1+k_2)x} \upd x\nonumber\\
    &=0,\label{orthogonality1}
\end{align}
since $\mu\omega_1\omega_2= \red{\mu c^2 k_1 k_2} = Tk_1k_2$. Second, we show the orthogonality of solutions travelling in the same direction
\begin{align}
        \left\langle\begin{bmatrix}
        \breve{u}_\pm\\ \upi \breve{v}_\pm
    \end{bmatrix}(\cdot,\omega_1),\begin{bmatrix}
        \breve{u}_\pm\\ \upi \breve{v}_\pm
    \end{bmatrix}(\cdot,\omega_2)\right\rangle&=\intinf\mu(-\upi\omega_1 e^{\pm\upi k_1x})(-\upi\omega_2 e^{\pm\upi k_2 x})^*\nonumber\\
    &\qquad+T(\pm\upi k_1 e^{\pm\upi k_1x})(\pm\upi k_2 e^{\pm\upi k_2x})^*\upd x\nonumber\\
    &= 2\mu\omega_1\omega_2\intinf e^{\pm\upi (k_1-k_2)x}\upd x\nonumber\\
    &=4\pi\mu\omega_1^2\delta(k_1-k_2)\nonumber\\
    &=4\pi \mu c \omega_1^2 \delta(\omega_1-\omega_2).\label{orthogonality2}
\end{align}
Since the frequency domain solutions exist for all $\omega$, the spectrum of $\mathscr{P}$ is the entire real line. As such, the spectral theorem allows us to write the full time domain solution as a continuous superposition of the eigenfunctions of $\mathscr{P}$, that is
\begin{equation}\label{general_time_domain_sol}
    \begin{bmatrix}
        u\\\upi v
    \end{bmatrix}(x,t)=\intinf C_+(t,\omega)\begin{bmatrix}
        \breve{u}_+\\ \upi \breve{v}_+
    \end{bmatrix}(x,\omega)+C_-(t,\omega)\begin{bmatrix}
        \breve{u}_-\\ \upi \breve{v}_-
    \end{bmatrix}(x,\omega)\upd \omega.
\end{equation}
Substituting \eqref{general_time_domain_sol} into \eqref{first_order_system} gives
\begin{align}
    \upi \partial_t\begin{bmatrix}
        u\\\upi v
    \end{bmatrix}(x,t)&=\upi\intinf \partial_t C_+(t,\omega)\begin{bmatrix}
        \breve{u}_+\\ \upi \breve{v}_+
    \end{bmatrix}(x,\omega)+\partial_t C_-(t,\omega)\begin{bmatrix}
        \breve{u}_-\\ \upi \breve{v}_-
    \end{bmatrix}(x,\omega)\upd \omega\nonumber\\
   &=\intinf C_+(t,\omega)\mathscr{P}\begin{bmatrix}
        \breve{u}_+\\ \upi \breve{v}_+
    \end{bmatrix}(x,\omega)+C_-(t,\omega)\mathscr{P}\begin{bmatrix}
        \breve{u}_-\\ \upi \breve{v}_-
    \end{bmatrix}(x,\omega)\upd \omega\nonumber\\
    &=\intinf C_+(t,\omega)\omega\begin{bmatrix}
        \breve{u}_+\\ \upi \breve{v}_+
    \end{bmatrix}(x,\omega)+C_-(t,\omega)\omega\begin{bmatrix}
        \breve{u}_-\\ \upi \breve{v}_-
    \end{bmatrix}(x,\omega)\upd \omega.
\end{align}
Equivalently, the above can be restated as
\begin{align}
    &\intinf (\upi\partial_t C_+(t,\omega)-\omega C_+(t,\omega))\begin{bmatrix}
        \breve{u}_+\\ \upi \breve{v}_+
    \end{bmatrix}(x,\omega)\\
    &\qquad+(\upi\partial_t C_-(t,\omega)-\omega C_-(t,\omega))\begin{bmatrix}
        \breve{u}_-\\ \upi \breve{v}_-
    \end{bmatrix}(x,\omega)\upd \omega=0.
\end{align}
After applying the orthogonality relations \eqref{orthogonality1} and \eqref{orthogonality2}, we obtain
\begin{equation}
    \upi \partial_t C_\pm(t,\omega)=\omega C_\pm(t,\omega)\quad\text{which implies}\quad C_\pm(t,\omega)=A_\pm(\omega)e^{-\upi \omega t}.
\end{equation}
Substituting this expression back into \eqref{general_time_domain_sol} gives
\begin{equation}\label{general_time_domain_sol2}
    \begin{bmatrix}
        u\\\upi v
    \end{bmatrix}(x,t)=\intinf \left(A_+(\omega)\begin{bmatrix}
        \breve{u}_+\\ \upi \breve{v}_+
    \end{bmatrix}(x,\omega)+A_-(\omega)\begin{bmatrix}
        \breve{u}_-\\ \upi \breve{v}_-
    \end{bmatrix}(x,\omega)\right)e^{-\upi\omega t}\upd \omega.
\end{equation}
The unknown spectral amplitudes $A_\pm$ are determined from the initial conditions \eqref{ic1}, \eqref{ic2}, i.e. they must satisfy
\begin{equation}\label{initial_condition_relation}
    \begin{bmatrix}
        f\\\upi g
    \end{bmatrix}(x)=\intinf A_+(\omega)\begin{bmatrix}
        \breve{u}_+\\ \upi \breve{v}_+
    \end{bmatrix}(x,\omega)+A_-(\omega)\begin{bmatrix}
        \breve{u}_-\\ \upi \breve{v}_-
    \end{bmatrix}(x,\omega)\upd \omega.
\end{equation}
\red{Taking the inner product on both sides of \eqref{initial_condition_relation} with the frequency-domain solution and using the orthogonality relations \eqref{orthogonality1} and \eqref{orthogonality2}, we obtain}
\begin{equation}
    \left\langle\begin{bmatrix}
        f\\ \upi g
    \end{bmatrix},\begin{bmatrix}
        \breve{u}_\pm\\ \upi v_\pm
    \end{bmatrix}(\cdot,\omega)\right\rangle=4\pi\mu c \omega^2A_\pm(\omega).
\end{equation}
We rearrange the above and use integration by parts, which yields
\begin{align}
    A_+(\omega)&=\frac{1}{4\pi \omega^2\mu c}\intinf T \partial_x f(x)(\partial_x u_+(x,\omega))^*+\mu g(x)(v_+(x,\omega))^*\mathrm{d}x\nonumber\\
    &=\frac{1}{4\pi \omega^2\mu c}\intinf -T f(x)(\partial_x^2 u_+(x,k))^*+\mu g(x)(v_+(x,k))^*\mathrm{d}x\nonumber\\
    &=\frac{1}{4\pi \omega^2\mu c}\intinf -T f(x)((-\upi k)^2e^{-\upi k x})+\mu g(x)(\upi \omega e^{-\upi k x})\mathrm{d}x\nonumber\\
    &=\frac{1}{4\pi c}\intinf f(x)e^{-\upi k x}\mathrm{d}x+\frac{\upi}{4\pi \omega c}\intinf g(x)e^{-\upi k x}\mathrm{d}x,\label{A_plus}
\end{align}
and equivalently
\begin{equation}
     A_-(\omega)=\frac{1}{4\pi c}\intinf f(x)e^{\upi k x}\mathrm{d}x+\frac{\upi}{4\pi \omega c}\intinf g(x)e^{\upi k x}\mathrm{d}x.\label{A_minus}
\end{equation}
Substituting \eqref{A_plus} and \eqref{A_minus} into \eqref{general_time_domain_sol2}, interchanging the order of integration and rearranging gives
\begin{align}
    u(x,t)&= \frac{1}{4\pi c}\intinf f(x^\prime) \intinf (e^{\upi \omega (x-x^\prime -ct)/c}+e^{\upi \omega (x^\prime-x -ct)/c})\mathrm{d}\omega\mathrm{d}x^\prime\nonumber\\
    &\quad+\frac{\upi}{4\pi c}\intinf g(x^\prime)\intinf\frac{1}{\omega}(e^{\upi \omega (x-x^\prime -ct)/c}+e^{\upi \omega (x^\prime-x -ct)/c})\mathrm{d}\omega\mathrm{d}x^\prime.
\end{align}
Evaluating the integrals with respect to $\omega$ \red{(in the sense of distributions)} then gives
\begin{align}
    u(x,t)&= \frac{1}{4\pi c}\intinf f(x^\prime) 2\pi (\delta((x-x^\prime-ct)/c)+\delta((x^\prime-x-ct)/c))\mathrm{d}x^\prime\nonumber\\
    &\quad+\frac{1}{4 c}\intinf g(x^\prime)(\sgn((x-x^\prime-ct)/c)+\sgn((x^\prime-x-ct)/c))\mathrm{d}x^\prime.
\end{align}
Note that
\begin{align}
    \sgn((x-x^\prime-ct)/c)+\sgn((x^\prime-x-ct)/c)&=\sgn(x-x^\prime-ct)+\sgn(x^\prime-x-ct)\nonumber\\
    &=\sgn(x^\prime-(x+ct))-\sgn(x^\prime-(x-ct))\nonumber\\
    &=\begin{cases}
        0&x^\prime<x-ct\\
        2&x-ct<x^\prime<x+ct\\
        0&x^\prime>x+ct
    \end{cases}\red{.}
\end{align}
Thus
\begin{align}
    u(x,t)&= \tfrac{1}{2}(f(x-ct)+f(x+ct))+\frac{1}{2c}\int_{x-ct}^{x+ct} g(x^\prime)\upd x^\prime,
\end{align}
which is precisely d'Alembert's formula for the one-dimensional wave equation.

\section{A string with a simple resonant scatterer}\label{mass-spring_sec}
\begin{figure}
    \centering
    \includegraphics[width=\textwidth]{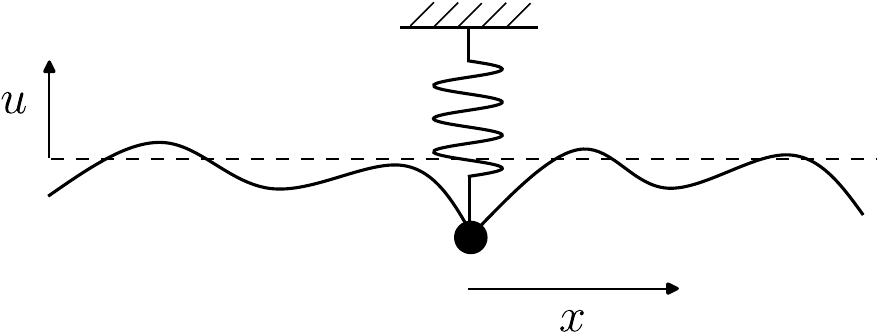}
    \caption{\red{A schematic of the mass-spring scattering problem.}}
    \label{fig:schematic}
\end{figure}

We now consider the case where a mass-spring system is coupled to the motion of the string, \red{which was first considered in \cite{lamb1900peculiarity}.} In particular, a point mass $M$ is bound to the string at $x=0$ and is also attached to a spring with spring constant $k_s$. \red{The other end of the string is fixed. A schematic of the problem is given in figure \ref{fig:schematic}.}

The kinematic and dynamic equations for the mass are
\begin{subequations}\label{time_domain_boundary_conditions}
    \begin{align}
    u(0^+,t)&=u(0^-,t)\label{kinematic}\\
    M\partial_t^2 u(0,t)&=T(\partial_x u(0^+,t)-\partial_x u(0^-,t))-k_s u(0,t),\label{dynamic}
\end{align}
\end{subequations}
and the motion of the string is governed by the wave equation \eqref{wave_equation} for $x\neq 0$. We again impose the initial conditions \eqref{ic1} and \eqref{ic2}.

\subsection{Solution using the GEM}
We seek a solution of the time-domain problem using the GEM. Analogously to \eqref{first_order_system} and \eqref{L_def}, the system can be restated as
\begin{equation}\label{two-component_td}
    \upi \partial_t\begin{bmatrix}u\\ \upi v\end{bmatrix}=\mathscr{P}\begin{bmatrix}u\\ \upi v\end{bmatrix},
\end{equation}
where
\begin{equation}
    \mathscr{P}=\begin{bmatrix}
        0&1\\ G&0
    \end{bmatrix}
\end{equation}
and
\begin{equation}
    Gu=\begin{cases}
        -\frac{T}{M}(\partial_x u(0^+,t)-\partial_x u(0^-,t))+\frac{k_s}{M}u(0,t)&x=0\\
        -c^2\partial_x^2 u&x\neq 0.
    \end{cases}
\end{equation}
The energy of the system is given by
\begin{equation}
    E=\tfrac{1}{2}\intinf\mu|v(x,t)|^2+T|\partial_x u(x,t)|^2\upd x+\tfrac{1}{2}M|v(0,t)|^2+\tfrac{1}{2}k_s|u(0,t)|^2,
\end{equation}
which motivates us to introduce an inner product of the form
\begin{equation}\label{inner_product_mass_spring}
    \left\langle\begin{bmatrix}
        u_1\\ \upi v_1
    \end{bmatrix},\begin{bmatrix}
        u_2\\ \upi v_2
    \end{bmatrix}\right\rangle = \intinf T(\partial_x u_1)(\partial_x u_2)^* + \mu v_1v_2^* \upd x+Mv_1(0)v_2(0)^*+k_su_1(0)u_2(0)^*.
\end{equation}
\red{The associated function space is}
\red{\begin{equation}
    \mathscr{H}=\left\{\left.\begin{bmatrix}
        u\\ \upi v
    \end{bmatrix}:\mathbb{R}\to\mathbb{C}^2\right|\left\langle\begin{bmatrix}
        u\\ \upi v
    \end{bmatrix},\begin{bmatrix}
        u\\ \upi v
    \end{bmatrix}\right\rangle<\infty\right\}.
\end{equation}
Note that this space is different from \eqref{fun_space_1} because the inner product has been redefined.} Following similar steps to \eqref{self-adjoint-proof} (that is, by direct computation using integration by parts), it is straightforward to show that the operator $\mathscr{P}$ is self-adjoint with respect to $\langle\cdot,\cdot\rangle$ and thus has real eigenvalues $\omega$. The corresponding generalised eigenfunctions are the solutions to the following frequency-domain problem:
\begin{subequations}\label{frequency_domain_problem}
\begin{align}
    \partial_x^2\breve{u}(x,\omega)&=-k^2\breve{u}(x,\omega),&x\neq 0\\
    \breve{u}(0^+,\omega)&=\breve{u}(0^-,\omega)\label{boundary_condition1}&\\
    \partial_x\breve{u}(0^+,\omega)-\partial_x\breve{u}(0^-,\omega)&=\frac{1}{T}(k_s-M\omega^2)\breve{u}(0,\omega)\label{boundary_condition2}\\
    \breve{v}(x,\omega)&=-\upi\omega\breve{u}(x,\omega).
\end{align}
\end{subequations}
We seek a solution to the frequency-domain scattering problem for an incident plane wave from $x=-\infty$ of the form
\begin{equation}\label{left_incident_FD}
    \breve{u}_+(x,\omega)=\begin{cases}
        e^{\upi k x}+r(\omega)e^{-\upi kx}&x<0\\
        t(\omega)e^{\upi kx}&x>0,
    \end{cases}
\end{equation}
where $r(\omega)$ and $t(\omega)$ are the reflection and transmission coefficient, respectively. It follows from the matching conditions \eqref{boundary_condition1} and \eqref{boundary_condition2} that
\begin{subequations}\label{r_and_t_coeff}
    \begin{align}
    r(\omega)&=\frac{k_s-M\omega^2}{2\upi k T-k_s+M\omega^2}\\
    t(\omega)&=\frac{2\upi k T}{2\upi k T-k_s+M\omega^2}.
\end{align}
\end{subequations}
The solution to the right-incident problem (i.e. for plane waves incoming from $x=+\infty$) can be obtained from the symmetry mapping $x\to-x$, which gives
\begin{equation}\label{right_incident_FD}
    \breve{u}_-(x,\omega)=
    \begin{cases}
        t(\omega)e^{-\upi kx}&x<0\\
        e^{-\upi k x}+r(\omega)e^{\upi kx}&x>0.
    \end{cases}
\end{equation}
As in \textsection\ref{dAlembert_sec}, the existence of the frequency domain solutions for all $\omega$ and the spectral theorem allows us to express the time-domain solution as a continuous superposition over the eigenfunctions
\begin{equation}\label{general_time_domain_sol3}
    \begin{bmatrix}
        u\\\upi v
    \end{bmatrix}(x,t)=\intinf \left(A_+(\omega)\begin{bmatrix}
        \breve{u}_+\\ \upi \breve{v}_+
    \end{bmatrix}(x,\omega)+A_-(\omega)\begin{bmatrix}
        \breve{u}_-\\ \upi \breve{v}_-
    \end{bmatrix}(x,\omega)\right)e^{-\upi\omega t}\upd \omega.
\end{equation}
Since \eqref{general_time_domain_sol3} must coincide with the initial conditions at $t=0$, the initial conditions determine the spectral amplitude functions $A_\pm$. To find these coefficients, we must compute the inner product of two frequency-domain solutions with the same direction of incidence at frequencies $\omega_1$ and $\omega_2$. We perform this calculation in appendix \ref{appendix_orthogonality} and obtain
\begin{equation}\label{same_normalisation}
    \left\langle\begin{bmatrix}
        \breve{u}_\pm\\ \upi v_\pm
    \end{bmatrix}(\cdot,\omega_1),\begin{bmatrix}
        \breve{u}_\pm\\ \upi v_\pm
    \end{bmatrix}(\cdot,\omega_2)\right\rangle=4\pi\mu c\omega_1\omega_2\delta(\omega_1-\omega_2).
\end{equation}
Thus the normalisation of the frequency-domain solutions is identical to the case in which there is no scatterer, which we derived in \eqref{orthogonality2}. This fact appears to hold generally \cite{povzner1953expansion,ikebe1960eigenfunction,wilcox1975scattering,hazard2002,hazard2007generalized}. From a physical perspective, this fact can be interpreted as the incident wave $e^{\upi k x}$ setting the energy for the problem irrespective of any scattering. After substituting \eqref{same_normalisation} into \eqref{general_time_domain_sol3} and setting $t=0$, we obtain
\begin{align}\label{A_pm_1}
    A_\pm(\omega)=&\frac{1}{4\pi\omega^2\mu c}\bigg[\intinf T\partial_xf(x^\prime)\partial_x\breve{u}_\pm(x^\prime,\omega)^*+\mu g(x^\prime)\breve{v}(x^\prime,\omega)^*\mathrm{d}x^\prime\nonumber\\
    &\quad+Mg(0)\breve{v}_\pm(0,\omega)^*+k_sf(0)\breve{u}_\pm(0,\omega)^*\bigg].
\end{align}
The first term in the integrand needs to be dealt with carefully due to the discontinuity of $\partial_x\breve{u}_\pm$ at $x=0$. We have
\begin{align}
    &\intinf T\partial_xf(x^\prime)\partial_x\breve{u}_\pm(x^\prime,\omega)^*\mathrm{d}x^\prime\nonumber\\
    &\quad=T \lim_{\epsilon\to 0}\left(\int_{-\infty}^{-\epsilon}+\int_{\epsilon}^\infty\right)\partial_xf(x^\prime)\partial_x\breve{u}_\pm(x^\prime,\omega)^*\mathrm{d}x^\prime\nonumber\\
    &\quad=T\lim_{\epsilon\to 0}\bigg[f(x^\prime)\partial_x\breve{u}_\pm(x^\prime,\omega)^*\bigg |_{-\infty}^{-\epsilon}+f(x^\prime)\partial_x\breve{u}_\pm(x^\prime,\omega)^*\bigg |_\epsilon^\infty\nonumber\\
    &\quad\quad-\left(\int_{-\infty}^{-\epsilon}+\int_{\epsilon}^{\infty}\right)f(x^\prime)\partial_x^2\breve{u}_\pm(x^\prime,\omega)^*\mathrm{d}x^\prime\bigg]\nonumber\\
    &\quad=Tf(0)(\partial_x\breve{u}(0^-,\omega)-\partial_x\breve{u}(0^+,\omega))^*+T\lim_{\epsilon\to 0}\left(\int_{-\infty}^{-\epsilon}+\int_{\epsilon}^{\infty}\right)f(x^\prime)k^2\breve{u}_\pm(x^\prime,\omega)^*\mathrm{d}x^\prime\nonumber\\
    &\quad=f(0)(M\omega^2-k_s)\breve{u}_\pm(0,\omega)^*+k^2T\intinf f(x^\prime)\breve{u}_\pm(x^\prime,\omega)^*\mathrm{d}x^\prime,\label{integral_with_limit}
\end{align}
where we have used \eqref{frequency_domain_problem}. Substituting \eqref{integral_with_limit} into \eqref{A_pm_1} then gives
\begin{align}
    A_\pm(\omega)=&\frac{1}{4\pi \mu c}\bigg[ Mf(0)\breve{u}_\pm(0,\omega)^*+\mu\intinf f(x^\prime)\breve{u}_\pm(x^\prime,\omega)^*\mathrm{d}x^\prime\nonumber\\
    &\quad+\frac{M}{\omega^2}g(0)\breve{v}_\pm(0,\omega)^*+\frac{\mu}{\omega^2}\intinf g(x^\prime)\breve{v}_\pm(x^\prime,\omega)^*\mathrm{d}x^\prime\bigg].\label{A_pm_2}
\end{align}
\red{Equation \eqref{A_pm_2} can be viewed as a transform from the spatial domain to the frequency domain, while \eqref{general_time_domain_sol3} is the corresponding inverse transform from the frequency domain to the spatial domain. In other words,  \eqref{A_pm_2} and \eqref{general_time_domain_sol3} form a transform/inverse transform pair. Because the mapping is described by the frequency-domain solutions, this transform is distinct from the Fourier transform.}

\subsection{Computation and results}
The GEM is well suited to computation, as it reduces to matrix multiplication after approximating the integrals using quadrature \cite{Meylan2023MWSW03}. In this subsection, we derive a matrix formula for the discretised GEM from \eqref{general_time_domain_sol3} and \eqref{A_pm_2}. Our derivation is analogous to the derivation of the discrete Fourier transform from the Fourier transform.

We discretise a bounded interval $[x_{\mathrm{min}},x_{\mathrm{max}}]$ of the spatial domain into $N_x$ subintervals of equal length $\Delta x$ with midpoints $x_j$. Let $j_0$ be such that $x_{j_0}=0$. We discretise a bounded frequency interval $[\omega_{\mathrm{min}},\omega_{\mathrm{max}}]$ into $N_\omega$ subintervals of width $\Delta\omega$ with midpoints $\omega_i$. We avoid having to solve the static problem (that is, the solution of \eqref{frequency_domain_problem} with $k=0$) by assuming $0\notin \{\omega_i\}$. Using midpoint quadrature, \eqref{A_pm_2} becomes 
\begin{align}
    A_\pm(\omega_i)=&\frac{1}{4\pi \mu c}\bigg[ M f(x_{j_0})\breve{u}_\pm(x_{j_0},\omega_i)^*+\mu\sum_{j=1}^{N_x}f(x_j)\breve{u}_\pm(x_j,\omega_i)^*\Delta x\nonumber\\
    &\quad+\frac{M}{\omega_i^2}g(x_{j_0})\breve{v}_\pm(x_{j_0},\omega)^*+\frac{\mu}{\omega_i^2}\sum_{j=1}^{N_x} g(x_j)\breve{v}_\pm(x_j,\omega_i)^*\Delta x\bigg],
\end{align}
or in matrix-vector notation
\begin{align}
    \mathbf{A}_\pm=&\frac{1}{4\pi \mu c}(\breve{\mathsfbi{U}}_\pm^\dag\mathbf{\Lambda}\mathbf{f}+ \mathrm{Diag}(1/\omega_i^2)\breve{\mathsfbi{V}}_\pm^\dag\mathbf{\Lambda}(\upi \mathbf{g})),
\end{align}
where $\mathbf{A}_\pm$ has entries $A_\pm(\omega_i)$ and $\breve{\mathsfbi{U}}_\pm$ and $\breve{\mathsfbi{V}}_\pm$ have entries $\breve{u}_\pm(x_j,\omega_i)$ and $\breve{v}_\pm(x_j,\omega_i)$, respectively. Moreover, $\dag$ denotes the conjugate transpose and $\mathbf{\Lambda}=\mathrm{Diag}(\mu\Delta x + \delta_{j, j_0})$. Lastly, $\mathbf{f}$ and $\mathbf{g}$ are vectors with the entries $f(x_j)$ and $g(x_j)$, respectively. After approximating the integral in \eqref{general_time_domain_sol3} with respect to $\omega$ using midpoint quadrature, the full time domain solution can be approximated as
\begin{equation}
    \mathbf{u}_t=\Delta\omega(\breve{\mathsfbi{U}}_+\mathsfbi{E}_t\mathbf{A}_++\breve{\mathsfbi{U}}_-\mathsfbi{E}_t\mathbf{A}_-)
\end{equation}
where $\mathbf{u}_t$ has entries $u(x_j,t)$ and $\mathsfbi{E}_t=\mathrm{Diag}(e^{-\upi \omega_i t})$.

\begin{figure}
    \centering
    \includegraphics[width=\textwidth]{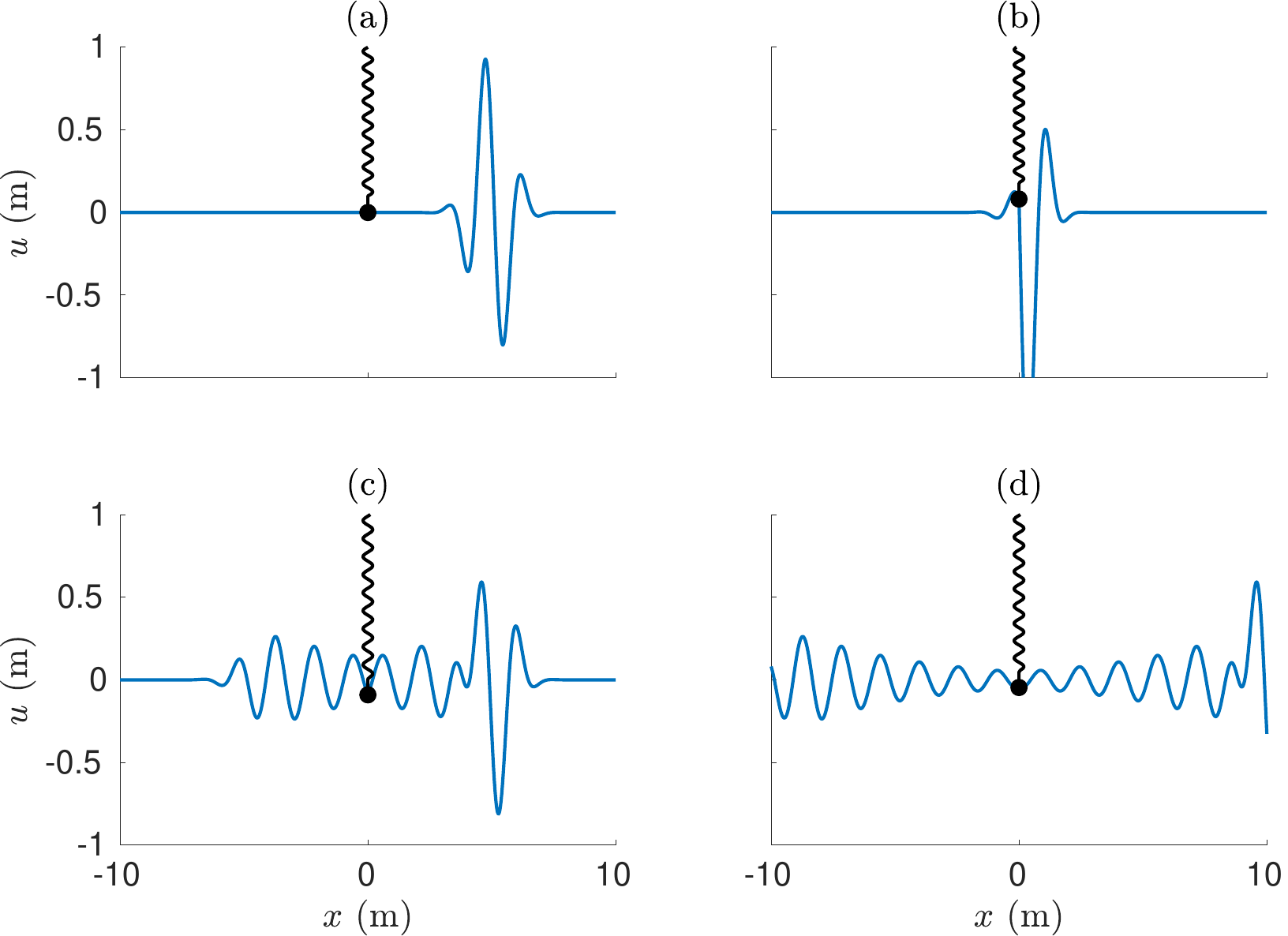}
    \caption{String displacement (blue line) induced by an incident wave packet given by $f(x)=e^{-(x-5)^2}\cos(4x)$ and $g(x)=cf^\prime(x)$ at (a) $t=0$\,s, (b) $t=5$\,s, (c) $t=10$\,s and (d) $t=15$\,s. The physical parameters are $\mu=1$\,kg\,m$^{-1}$, $T=1$\,N, $k_s=80$\,kg\,s$^{-2}$ and $M=5$\,kg. The numerical parameters used to generate the figure are $[x_{\mathrm{min}},x_{\mathrm{max}}]=[-10,10]$\,m, $\Delta x=0.025$\,m\red{,} $[\omega_{\mathrm{min}},\omega_{\mathrm{max}}]=[-20,20]$\,s$^{-1}$ and $\Delta\omega=40/799$\,s$^{-1}$. The mass-spring system is shown symbolically. Note that animations corresponding to all figures in this paper are provided in the supplementary material.}
    \label{fig1}
\end{figure}

Note that other quadrature methods (e.g. Simpson's rule) could be used in place of the midpoint rule used here. However, the frequency-domain solution matrices $\breve{\mathsfbi{U}}_\pm$ and $\breve{\mathsfbi{V}}_\pm$ are computationally inexpensive due to the simple nature of the current problem. As such, we can simply reduce $\Delta x$ and $\Delta \omega$ whenever more accuracy is required. Here, accuracy is determined visually by comparing the GEM solution at $t=0$\,s with the initial conditions. Figure \ref{fig1} shows the solution for one particular choice of initial conditions. \red{The MATLAB code used to generate this result is provided in the supplementary file \texttt{mass\_spring\_code.mlx}.}

\section{The Singularity Expansion Method}\label{SEM_sec}
\subsection{Derivation}
\red{In this section, we derive the SEM and apply it to the problem considered in \textsection\ref{mass-spring_sec}. To begin, we take the Fourier transform in time of \eqref{two-component_td} to obtain
\begin{equation}\label{inhomogeneous_helmholtz}
    \upi (\mathscr{P}-\omega)\begin{bmatrix}
        \widehat{u}\\ \upi \widehat{v}
    \end{bmatrix}(x,\omega)=\mathbf{f}(x),
\end{equation}
where $[\widehat{u},\upi \widehat{v}]^\intercal$ is the Fourier transform of the solution $[u,\upi v]^\intercal$ with respect to time, and} we have defined $\mathbf{f}=[f,\upi g]^\intercal$. To enforce a unique solution of this equation, we also impose the Sommerfeld radiation condition
\begin{equation}\label{sommerfeld}
    (\partial_x\mp\upi k)\widehat{u}(x,\omega)\to 0\quad\text{as}\quad x\to\pm\infty.
\end{equation}
Equations \eqref{inhomogeneous_helmholtz} and \eqref{sommerfeld} can be solved for all $\omega\in\mathbb{C}$ except at the complex resonant frequencies $\omega_j$\red{, for which} there exist non-trivial two-component eigenfunctions $\boldsymbol{\Phi}_j=[\phi_j,\upi(-\upi\omega_j\phi_j)]^\intercal$ which satisfy
\begin{equation}
\mathscr{P}\boldsymbol{\Phi}_j=\omega_j\boldsymbol{\Phi}_j
\end{equation}
and \eqref{sommerfeld}, i.e. $(\partial_x\mp\upi k_j)\phi_j(x)\to 0$ as $x\to\pm\infty$, where $k_j=\omega_j/c$. Thus $\boldsymbol{\Phi}_j$ is a mode consisting of purely outgoing waves at a single complex frequency $\omega_j$.

In order to apply the SEM, we rearrange \eqref{inhomogeneous_helmholtz} and apply the inverse Fourier transform, which gives
\begin{equation}\label{inverse_FT_sol}
    \begin{bmatrix}
        u\\ \upi v
    \end{bmatrix}(x,t)=\frac{-\upi}{2\pi}\intinf e^{-\upi\omega t}(\mathscr{P}-\omega)^{-1}\mathbf{f}(x) \upd\omega.
\end{equation}
By assuming that the integrand is meromorphic in the lower half plane with poles at the complex resonant frequencies $\omega_j$, the residue theorem then gives
\begin{align}\label{residue_sum}
    \begin{bmatrix}
        u\\ \upi v
    \end{bmatrix}&=-\sum_j e^{-\upi\omega_j t}\mathrm{Res}\left((\mathscr{P}-\omega)^{-1}\mathbf{f}(x),\omega_j\right)+\frac{\upi}{2\pi}\lim_{R\to\infty}\int_{\gamma_R} e^{-\upi\omega t}(\mathscr{P}-\omega)^{-1}\mathbf{f}(x) \upd\omega,
\end{align}
where $\sum_j$ denotes the sum over all complex resonant frequencies $\omega_j$ and
\begin{equation}
    \gamma_R = \{R e^{-\upi s}|0\leq s\leq \pi\}.
\end{equation}
In order to apply the SEM approximation, we assume that the factor $e^{-\upi\omega t}$ causes the integral in \eqref{residue_sum} to vanish as $R\to \infty$ for sufficiently large $t$. To compute the residues in \eqref{residue_sum}, we use a result due to Steinberg for analytic families of compact operators \cite{steinberg1968meromorphic}. A simplified interpretation of this result in terms of analytic families of matrices was provided by Meylan and Tomic \cite{meylan2012complex}. For a simple pole at $\omega_j$, this result gives
\begin{subequations}\label{residue_formula}
    \begin{align}
    \mathrm{Res}\left((\mathscr{P}-\omega)^{-1}\mathbf{f}(x),\omega_j\right)&=\frac{\left\langle\mathbf{f},\boldsymbol{\Psi}_j\right\rangle}{\left\langle\left[\frac{\upd}{\upd\omega}(\mathscr{P}-\omega)\right]_{\omega=\omega_j}\boldsymbol{\Phi}_j,\boldsymbol{\Psi}_j\right\rangle}\boldsymbol{\Phi}_j(x)\\
    &=-\frac{\left\langle\mathbf{f},\boldsymbol{\Psi}_j\right\rangle}{\left\langle\boldsymbol{\Phi}_j,\boldsymbol{\Psi}_j\right\rangle}\boldsymbol{\Phi}_j(x).
\end{align}
\end{subequations}
Although we only encounter simple poles in this paper, note that \eqref{residue_formula} can be extended for poles of order greater than one \cite{meylan2009time}. In \eqref{residue_formula}, $\boldsymbol{\Psi}_j=[\psi_j,\upi(-\upi\omega_j^*\psi_j)]^\intercal$ is a non-trivial eigenvector of the adjoint of $\mathscr{P}-\omega_j$. It therefore satisfies
\begin{equation}
\mathscr{P}\boldsymbol{\Psi}_j=\omega_j^*\boldsymbol{\Psi}_j,
\end{equation}
as well as the \red{conjugate} of the Sommerfeld radiation condition \eqref{sommerfeld}, i.e.
\begin{equation}
    (\partial_x\pm\upi k_j^*)\psi_j(x)\to 0\quad\text{as}\quad x\to\pm\infty.
\end{equation}
Thus, $\boldsymbol{\Psi}_j$ is a mode consisting of purely incoming waves at a single complex frequency $\omega_j^*$. In contrast with the \textit{resonant mode} $\boldsymbol{\Phi}_j$, which consists of purely outgoing waves, we call $\boldsymbol{\Psi}_j$ the corresponding \textit{absorbing mode}. To complete the derivation of the singularity expansion formula, we substitute \eqref{residue_formula} into \eqref{residue_sum} giving
\begin{equation}\label{SEM_formula_1}
    \begin{bmatrix}
        u\\ \upi v
    \end{bmatrix}(x,t)\approx\sum_j e^{-\upi\omega_j t}\frac{\left\langle\mathbf{f},\boldsymbol{\Psi}_j\right\rangle}{\langle\boldsymbol{\Phi}_j,\boldsymbol{\Psi}_j\rangle}\boldsymbol{\Phi}_j(x).
\end{equation}
In the next subsection, we will apply the SEM to the problem of wave scattering by the mass-spring system, which we introduced in \textsection\ref{mass-spring_sec}.

\subsection{Computation and results}
The problem of scattering by a mass-spring system has only two complex resonances, which occur at the poles of the reflection (or transmission) coefficient. By factoring \red{the} denominator in \eqref{r_and_t_coeff}, we find that they are
\begin{equation}
    \omega_{1,2} = \frac{-iT}{cM}\pm\frac{1}{M}\sqrt{k_sM-\mu T}.\label{poles_of_reflection}
\end{equation}
Thus \eqref{SEM_formula_1} becomes
\begin{equation}
    u_s(x,t)\approx\sum_{j=1}^2 \frac{\langle\mathbf{f},\boldsymbol{\Psi}_j\rangle}{\langle\boldsymbol{\Phi}_j,\boldsymbol{\Psi}_j\rangle}\Phi(x)e^{-\upi\omega_j t}.\label{SEM}
\end{equation}
\red{It remains to find the resonant modes $\boldsymbol{\Phi}_j$ and absorbing modes $\boldsymbol{\Psi}_j$, for $j\in\{1,2\}$. Since the resonant modes are eigenvectors of $\mathscr{P}$, we have}
\begin{equation}
\boldsymbol{\Phi}_j=\begin{bmatrix}
        \phi_j\\ \upi (-\upi \omega_j \phi_j)
\end{bmatrix},
\end{equation}
where $\phi_j$ is the non-trivial solution to \eqref{frequency_domain_problem} at $k=k_j$ with no incoming wave given by
    \begin{equation}
        \phi_j(x)=\begin{cases}
            e^{-\upi k_jx}&x<0\\
            e^{\upi k_jx}&x>0.
        \end{cases}
    \end{equation}
\red{where $k_j=\omega_j/c$. Analogously, the absorbing modes are eigenvectors of $\mathscr{P}$ that correspond to the eigenvalues $\omega_j^*$ for $j\in\{1,2\}$, so that}
\begin{equation}
    \boldsymbol{\Psi}_j=\begin{bmatrix}
        \psi_j\\ \upi (-\upi \omega_j^* \psi_j)
    \end{bmatrix},
\end{equation}
where $\psi_j$ is a non-trivial solutions to \eqref{frequency_domain_problem} at $k=k_j^*$ with no outgoing wave given by
    \begin{equation}
        \psi_j(x)=\begin{cases}
            e^{\upi k_j^*x}&x<0\\
            e^{-\upi k_j^*x}&x>0.
        \end{cases}
    \end{equation}
Note that $\psi_j(x)=\phi_j(x)^*$. 

In general, the numerator of \eqref{SEM} (i.e. the inner product of the initial conditions with the absorbing mode) should be computed numerically. However, we can compute the denominator of \eqref{SEM} (i.e. the normalisation of the complex resonances) directly. We have
\begin{subequations}
    \begin{align}
    \langle\boldsymbol{\Phi}_j,\boldsymbol{\Psi}_j\rangle&=\intinf T\partial_x \phi_j(x) (\partial_x \psi_j(x))^*+\mu (-\upi\omega_j\phi_j(x))(-\upi\omega_j^*\psi_j(x))^*\upd x\nonumber\\
    &\quad+M (-\upi\omega_j\phi_j(0))(-\upi\omega_j^*\psi_j(0))^*+k_s\phi_j(0)\psi_j(0)^*\nonumber\\
    &=2\int_{-\infty}^0 T(-\upi k_je^{-\upi k_jx})(\upi k_j^*e^{\upi k_j^*x})^*+\mu (-\upi\omega_je^{-\upi k_jx})(-\upi\omega_j^*e^{\upi k_j^*x})^*\upd x\nonumber\\
    &\quad+M \omega_j^2+k_s\nonumber\\
    &=2\int_{-\infty}^0 \mu\omega_j^2e^{-2\upi k_jx}-Tk_j^2e^{-2\upi k_jx}\upd x+M \omega_j^2+k_s\label{SEM_normalisation_indeterminate}\\
    &=M \omega_j^2+k_s.\label{SEM_normalisation}
\end{align}
\end{subequations}
In \eqref{SEM_normalisation_indeterminate}, we have used the fact that $Tk_j^2=\mu\omega_j^2$ in order to eliminate the term containing the integral. This step is usually not mathematically valid because $\int_{-\infty}^0 e^{-2\upi k_jx}\upd x$ is divergent (since $k_j$ is in the lower half of the complex plane). However, it is known that the normalisation of the resonant modes can be computed by first regularising the divergent integrals \cite{Kristensen2020}. To this end, we assume that the formula  
\begin{equation}
    \int_{-\infty}^0 e^{ax}\upd x=\frac{1}{a},\label{analytic_continuation_int}
\end{equation}
which is only valid for $\Re\{a\}>0$, can be analytically continued into the region $\Re\{a\}<0$. It is in this sense that the formula obtained in \eqref{SEM_normalisation} holds. The vanishing of the semi-infinite parts of the normalisation integrals for the complex resonances of one-dimensional problems was previously observed by \cite{Kristensen2020} in the context of electromagnetic \red{wave} scattering by a dielectric barrier.

\begin{figure}
    \centering
    \includegraphics[width=\textwidth]{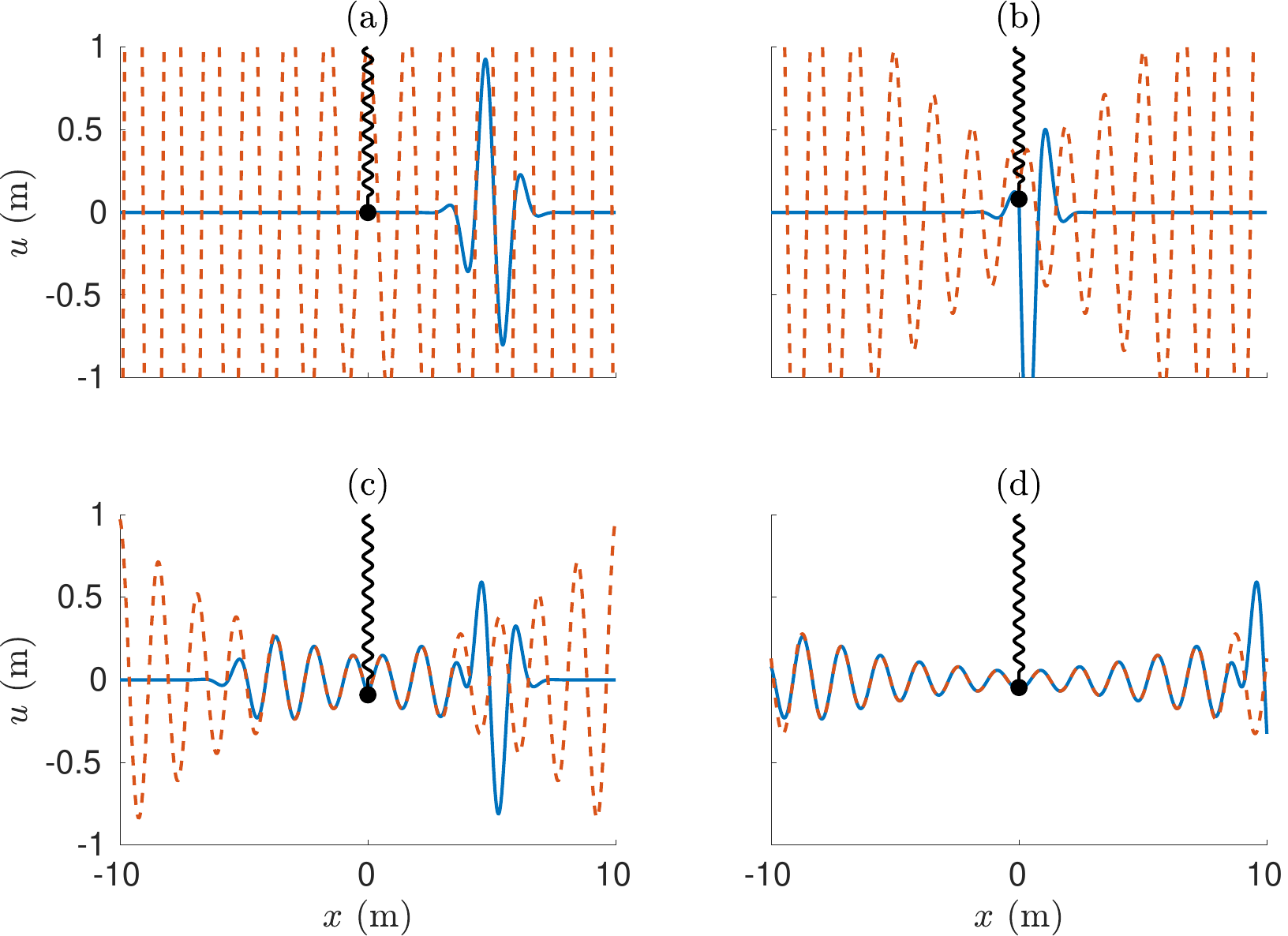}
    \caption{String displacement computed using the GEM (solid blue line) and SEM approximation (dashed red line) at (a) $t=0$\,s, (b) $t=5$\,s, (c) $t=10$\,s and (d) $t=15$\,s. The incident wave packet and physical parameters are identical to those in figure \ref{fig1}.}
    \label{fig:fig2}
\end{figure}

\begin{figure}
    \centering
    \includegraphics[width=\textwidth]{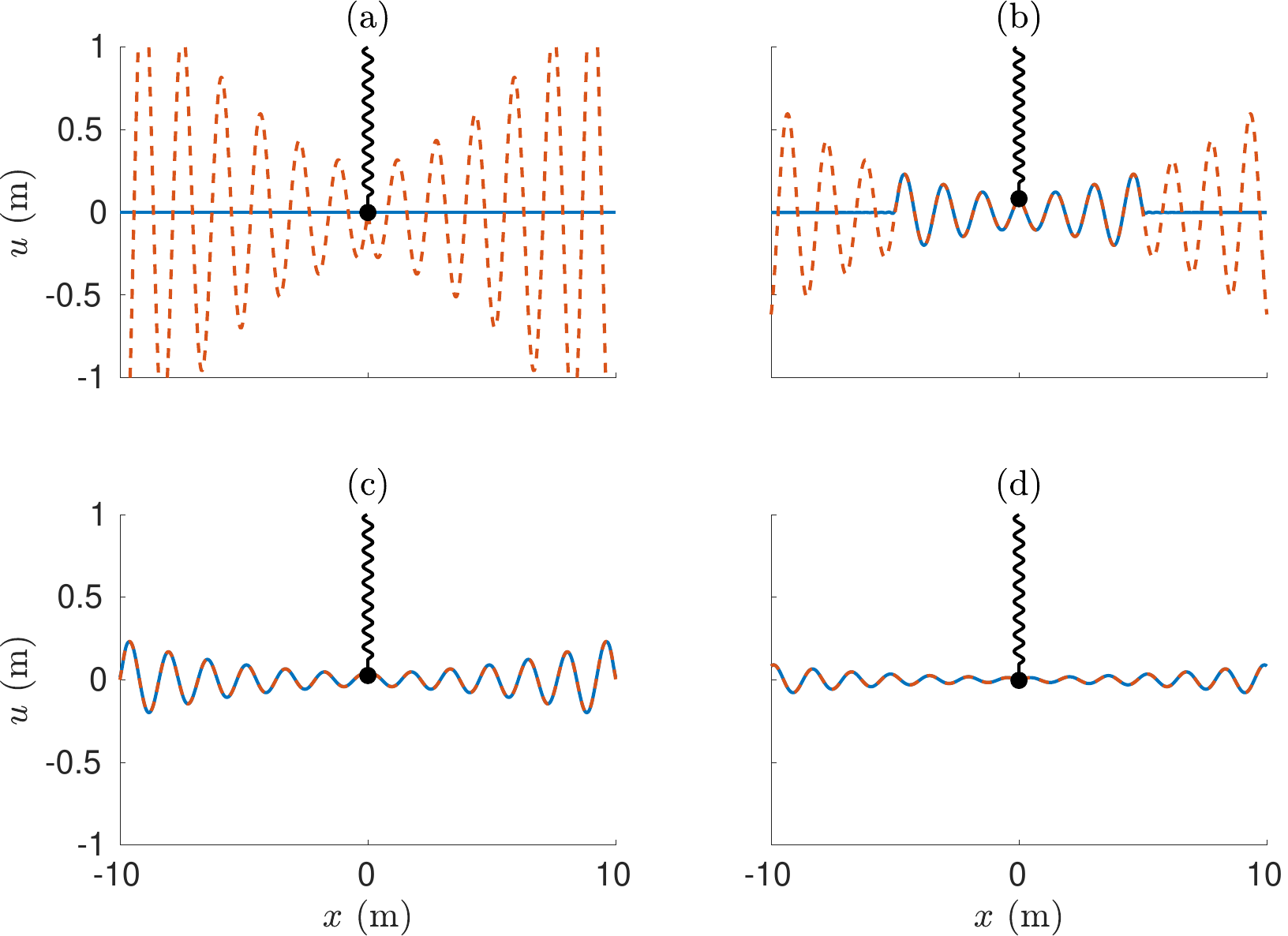}
    \caption{String displacement computed using the GEM (solid blue line) and SEM approximation (dashed red line) at (a) $t=0$\,s, (b) $t=5$\,s, (c) $t=10$\,s and (d) $t=15$\,s. The wave motion is induced by an impulse of the form $f(x)=0$, $g(0)=1$ and $g(x)=0$ for $x\neq 0$. The physical parameters are identical to those in figure \ref{fig1}.}
    \label{fig:fig3}
\end{figure}

Figures \ref{fig:fig2} and \ref{fig:fig3} compare the string displacement $u$ calculated using the SEM approximation and the GEM solution for two different choices of initial conditions. \red{The MATLAB code used to generate these results is provided in the supplementary file \texttt{mass\_spring\_code.mlx}.} In figure \ref{fig:fig2}, the SEM solution is inaccurate until some time $t_0$, which is approximately when the incident wave packet collides with the scatterer. After this, it qualitatively appears to agree with the GEM solution on the interval $(-c(t-t_0),c(t-t_0))$. Figure \ref{fig:fig3} compares the methods for an impulse-like initial condition on the mass. In this case, $t_0=0$ and the SEM describes the displacement of the mass $u(0)$ for all $t\geq 0$ (i.e. $t_0=0$). The accurate representation of the evolution of initial conditions which are only non-zero inside the resonator is a known feature of the SEM, which applies when the underlying structure of the problem meets the requirements of Lax-Phillips scattering theory (see \cite{pavlov1999irreversibility,meylan_2002} for further details).

\section{A mass on an infinite string next to an anchor point}\label{Anchor_point_sec}
\subsection{Preliminaries}
We consider the problem of scattering of waves on a semi-infinite stretched string by a mass at $x=0$. The string occupies the region $x>-L$ and is anchored in place at $x=-L$. The initial boundary value problem is
\begin{subequations}
    \begin{align}
\partial_t^2 u-c^2\partial_x^2 u&=0\\
u(-L,t)&=0\\
u(0^+,t)&=u(0^-,t)\label{mass_wall_matching_1}\\
M\partial_t^2 u(0,t)&=T(\partial_x u(0^+,t)-\partial_x u(0^-,t))\label{mass_wall_matching2}\\
u(x,0)&=f(x)\\
\partial_t u(x,0)&=g(x),
\end{align}
\end{subequations}
\red{As in \textsection\textsection\ref{dAlembert_sec} and \ref{mass-spring_sec}, we} write the problem as a two component system of the form
\begin{equation}
    \upi \partial_t\begin{bmatrix}u\\ \upi v\end{bmatrix}=\begin{bmatrix}
        0&1\\ G&0
    \end{bmatrix}\begin{bmatrix}u\\ \upi v\end{bmatrix}=
    \mathscr{P}\begin{bmatrix}u\\ \upi v\end{bmatrix},
\end{equation}
where
\begin{equation}
    \mathscr{P}\coloneqq\begin{bmatrix}
        0&1\\ G&0
    \end{bmatrix}
\end{equation}
and
\begin{equation}
    Gu\coloneqq\begin{cases}
        -\frac{T}{M}(\partial_x u(0^+,t)-\partial_x u(0^-,t))&x=0\\
        -c^2\partial_x^2 u&x\neq 0.
    \end{cases}
\end{equation}
The associated frequency-domain problem is
\begin{subequations}
    \begin{align}
    \partial_x^2\breve{u}(x,\omega)&=-k^2\breve{u}(x,\omega)\\
    \breve{u}(0^+,\omega)&=\breve{u}(0^-,\omega)\\
    \partial_x\breve{u}(0^+,\omega)-\partial_x\breve{u}(0^-,\omega)&=-\frac{M\omega^2}{T}\breve{u}(0,\omega)\\
    \breve{u}(-L,\omega)&=0
\end{align}
\end{subequations}
where $k=\omega/c$. The solution of the frequency-domain problem \red{to plane wave forcing} is
\begin{equation}
    \breve{u}(x,\omega) = \begin{cases}
        B\frac{\sin(k(x+L))}{\sin(kL)}&x<0\\
        e^{-\upi kx}+re^{\upi kx}&x>0,
    \end{cases}
\end{equation}
where the reflection coefficient $r$ and the interior amplitude $B$ are determined from the matching conditions \red{at $x=0$. We obtain}
\begin{subequations}
    \begin{align}
    r&=-\frac{Mc^2k^2-kT\cot(kL)-\upi kT}{Mc^2k^2-kT\cot(kL)+\upi kT}\label{reflection_coeff}\\
    B&=r+1.
\end{align}
\end{subequations}
Analogously to \eqref{inner_product_mass_spring}, we choose an energy inner product of the form
\begin{equation}
    \left\langle\begin{bmatrix}
 u_1\\ \upi v_1
    \end{bmatrix},\begin{bmatrix}
        u_2\\ \upi v_2
    \end{bmatrix}\right\rangle = \intL T(\partial_x u_1)(\partial_x u_2)^* + \mu v_1v_2^* \upd x+Mv_1(0)v_2(0)^*,
\end{equation}
\red{and associated function space}
\red{\begin{equation}
    \mathscr{H}=\left\{\left.\begin{bmatrix}
        u\\ \upi v
    \end{bmatrix}:[-L,\infty)\to\mathbb{C}^2\right|\left\langle\begin{bmatrix}
        u\\ \upi v
    \end{bmatrix},\begin{bmatrix}
        u\\ \upi v
    \end{bmatrix}\right\rangle<\infty\quad\text{and}\quad\partial_x u\biggr\rvert_{x=-L}=0\right\},
\end{equation}}
so that the operator $\mathscr{P}$ is self-adjoint. Following \textsection\ref{mass-spring_sec}, we compute the spectral amplitudes using the GEM as
\begin{subequations}
    \begin{align}
    A(\omega)&=\frac{1}{4\pi\mu c\omega^2}\left\langle\begin{bmatrix}
        f\\\upi g
    \end{bmatrix},\begin{bmatrix}
        \breve{u}\\\upi \breve{v}
    \end{bmatrix}(\cdot,\omega)\right\rangle\\
    &=\frac{1}{4\pi\mu c}\bigg[Mf(0)\breve{u}(0,\omega)^*+\mu\intL f(x)\breve{u}(x,\omega)^*\upd x\nonumber\\
    &\qquad+\frac{M}{\omega^2}g(0)\breve{v}(0,\omega)^*+\frac{\mu}{\omega^2}\intL g(x)\breve{v}(x,\omega)^*\upd x\bigg]
\end{align}
\end{subequations}
The generalised eigenfunction expansion for the time-domain solution can then be stated as
\begin{equation}
    u(x,t)=\intinf A(\omega)\breve{u}(x,\omega)e^{-\upi\omega t}\upd \omega.
\end{equation}
Alternatively, the SEM gives
\begin{equation}\label{SEM_full_expansion}
    u(x,t)\approx\sum_{j=-\infty}^\infty \frac{\langle\mathbf{f},\boldsymbol{\Psi}_j\rangle}{\langle\boldsymbol{\Phi}_j,\boldsymbol{\Psi}_j\rangle}\phi(x)e^{-\upi\omega_j t}.
\end{equation}
Note that in contrast to \textsection\ref{mass-spring_sec}, this problem supports infinitely many complex resonances as can be seen from the transcendental denominator of the reflection coefficient \eqref{reflection_coeff}.

\subsection{Locating the poles}
By rearranging the denominator of the reflection coefficient \eqref{reflection_coeff}, the complex resonant wavenumbers must satisfy the following transcendental equation
\begin{equation}
    \left(\frac{Mk}{\mu}+\upi\right)\sin(kL)-\cos(kL)=0.\label{transcendental1}
\end{equation}
For large $Mk/\mu$, the solutions $k_j$ are well approximated by $k_j\approx j\pi/L$. For $j\neq 0$, these turn out to be suitable initial guesses for numerical refinement using multiple iterations of Newton's method. However, this formula for the initial guess cannot be used to find $k_0$, since Newton's method fails to converge from the initial guess $0$\,m$^{-1}$. Instead, we seek $k_0$ by expanding the trigonometric functions in \eqref{transcendental1} in their Taylor series about $z=0$, where $z=kL$, which gives
\begin{equation}
    \left(\frac{M}{\mu L}z+\upi\right)\left(z-\frac{z^3}{6}+\dots\right)-\left(1-\frac{z^2}{2}+\dots\right)=0.\label{transcendental2}
\end{equation}
The estimate for $k_0$ is found by taking terms in \eqref{transcendental2} up to $O(z^2)$ and choosing the solution with a positive real part, which gives
\begin{equation}
    k_0\approx \frac{-\upi\mu+\mu\sqrt{\frac{4M}{\mu L}+1}}{2M+\mu L}
\end{equation}
This initial guess is refined using Newton's method to determine  $k_0$ accurately. Lastly, we note that $k_{-j}=k_{j-1}$ so only $k_j$ for $j\geq 0$ must be found numerically.

\subsection{Absorbing and resonant modes}
The absorbing modes are non-trivial solutions to the frequency-domain problem at $\omega_j^*$ with no outgoing wave, i.e.
\begin{equation}
    \psi_j(x) = \begin{cases}
        \frac{\sin(k_j^*(x+L))}{\sin(k_j^*L)}&x<0\\
        e^{-\upi k_j^*x}&x>0.
    \end{cases}
\end{equation}
The resonant modes are non-trivial solutions to the frequency-domain problem at $\omega_j$ with no incoming wave, i.e.
\begin{equation}
    \phi_j(x) = \begin{cases}
        \frac{\sin(k_j(x+L))}{\sin(k_jL)}&x<0\\
        e^{\upi k_jx}&x>0.
    \end{cases}
\end{equation}
When computing $\langle\boldsymbol{\Phi}_j,\boldsymbol{\Psi}_j\rangle$, we eliminate the integral over the semi-infinite region $(0,\infty)$, which follows from \eqref{SEM_normalisation} and subsequent discussion. The remaining component of the integral can be computed directly, i.e.
\begin{subequations}
    \begin{align}
\langle\boldsymbol{\Phi}_j,\boldsymbol{\Psi}_j\rangle&=M\omega_j^2+\frac{1}{\sin(k_jL)\sin^*(k_j^*L)}\int_{-L}^0\bigg(Tk_j^2\cos(k_j(x+L))\cos^*(k_j^*(x+L))\nonumber\\
    &\qquad +\mu\omega_j^2\sin(k_j(x+L))\sin^*(k_j^*(x+L))\bigg)\upd x\\
    &=M\omega_j^2+\frac{\mu\omega_j^2L}{\sin^2(k_jL)},
\end{align}
\end{subequations}
where we have used the Pythagorean identity ($\sin^2(z)+\cos^2(z)=1$) and the facts that $\sin^*(z^*)=\sin(z)$ and $\cos^*(z^*)=\cos(z)$.

\subsection{Results}

\begin{figure}
    \centering
    \includegraphics[width=\textwidth]{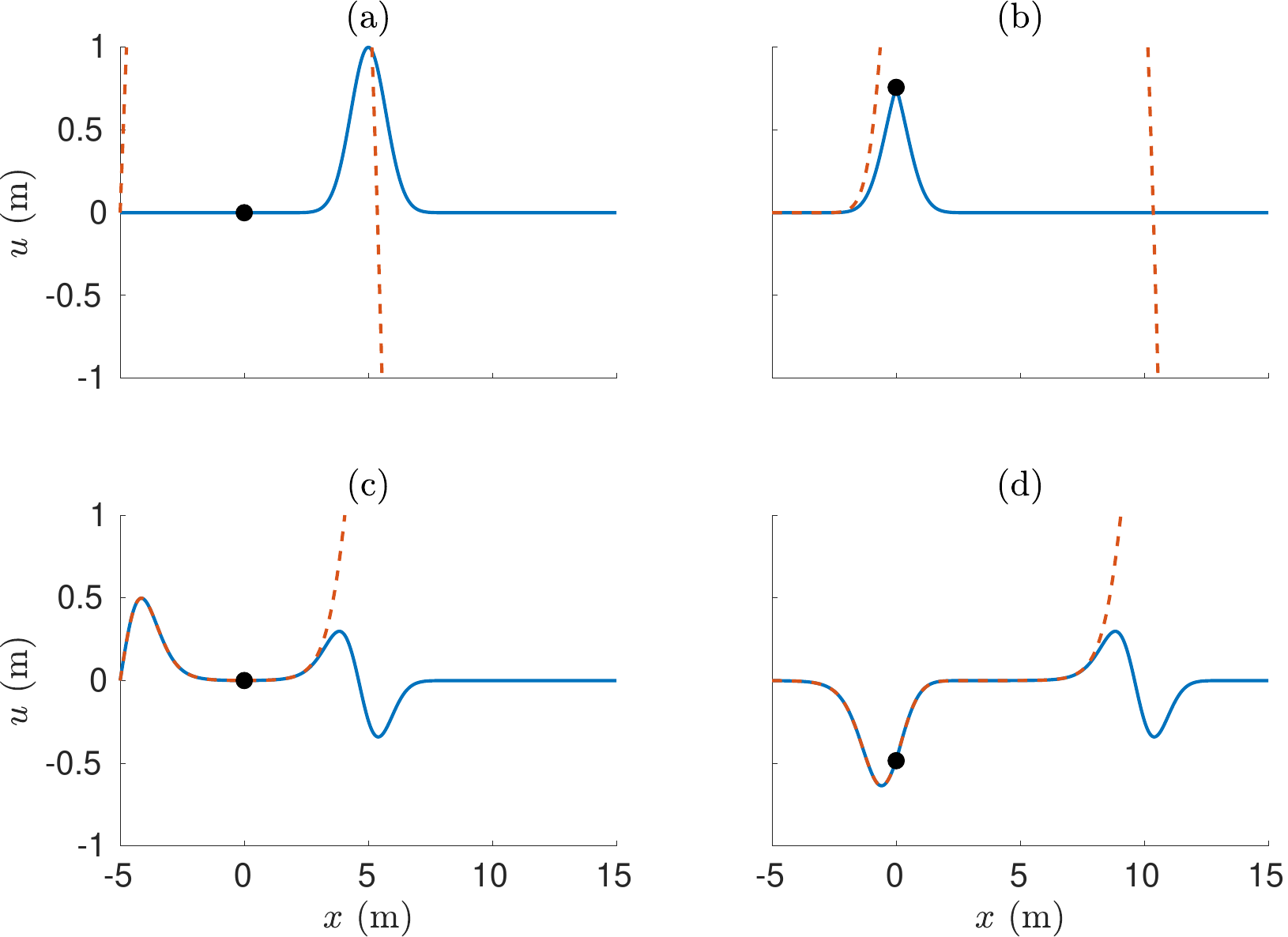}
    \caption{String displacement calculated using the GEM (solid blue line) and SEM approximation (dashed red line) at (a) $t=0$\,s, (b) $t=5$\,s, (c) $t=10$\,s and (d) $t=15$\,s. Wave motion is induced by an incident wave packet given by $f(x)=e^{-(x-5)^2}$ and $g(x)=cf^\prime(x)$. The physical parameters are $\mu=1$\,kg\,m$^{-1}$, $T=1$\,N, $L=5$\,m and $M=1$\,kg. The numerical parameters used to compute the GEM solution are $[x_{\mathrm{min}},x_{\mathrm{max}}]=[-5,15]$\,m, $\Delta x=0.025$\,m $[\omega_{\mathrm{min}},\omega_{\mathrm{max}}]=[-20,20]$\,s$^{-1}$ and $\Delta\omega=0.02$\,s$^{-1}$. To compute the SEM approximation, the sum in \eqref{SEM_full_expansion} was truncated to only contain the terms $-9\leq j \leq 8$. The mass is shown symbolically.}
    \label{fig4}
\end{figure}

\begin{figure}
    \centering
    \includegraphics[width=\textwidth]{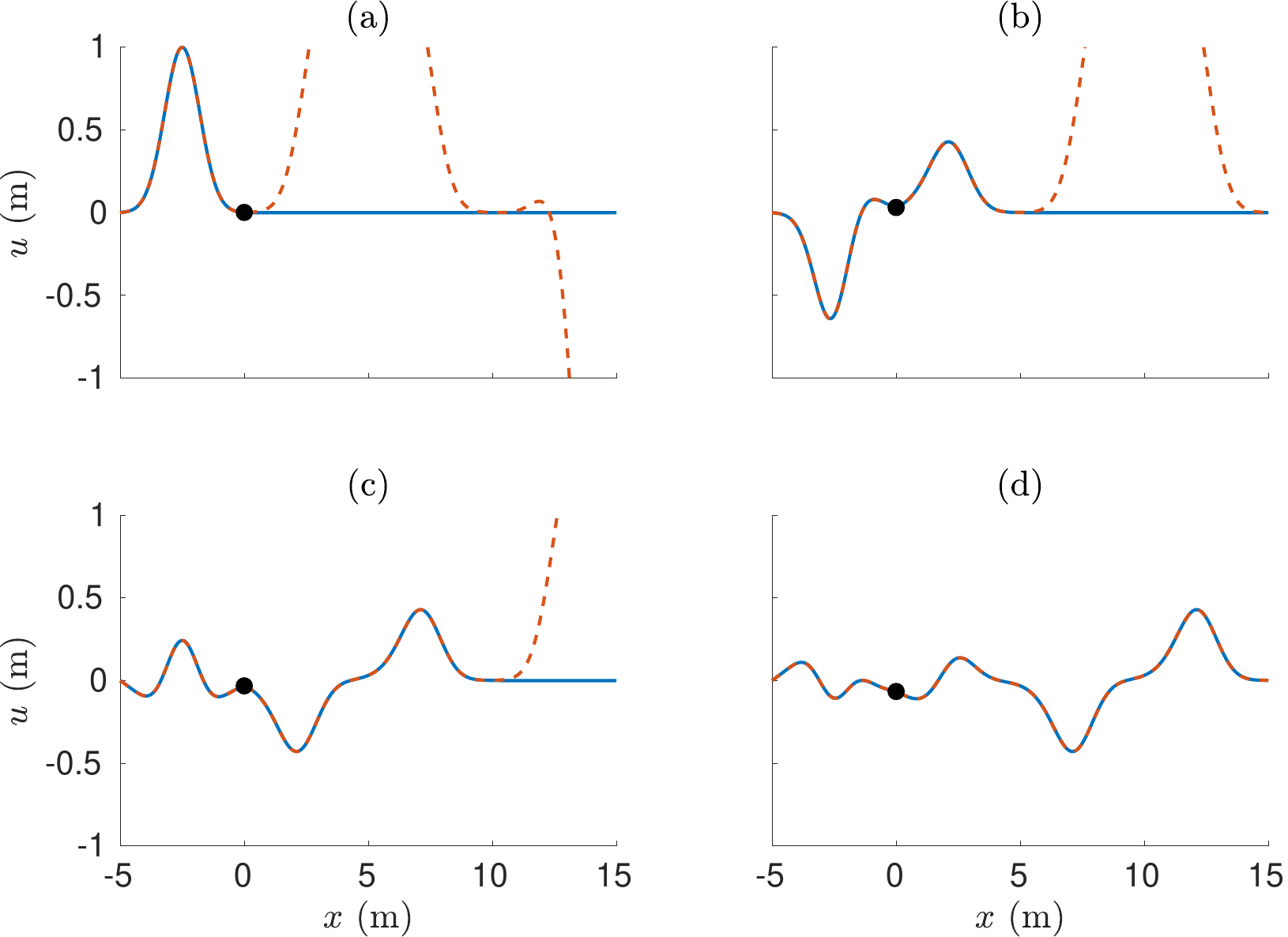}
    \caption{As for figure \ref{fig4} with the initial conditions $f(x)=e^{-(x+2.5)^2}$ and $g(x)=0$.}
    \label{fig5}
\end{figure}

Figures \ref{fig4} and \ref{fig5} show the solution to this problem using the GEM and SEM. In particular, figure \ref{fig4} shows the case of an incident Gaussian wave packet, while figure \ref{fig5} shows the case where a Gaussian initial condition is excited inside the resonator. In the former case, the SEM becomes accurate after the wave packet interacts with the resonator, whereas in the latter case, the SEM is accurate inside the resonator at all times. As we remarked in \textsection\ref{SEM_sec}, this is because the problem satisfies the requirements of Lax-Phillips scattering theory. The accuracy of the SEM is achieved despite truncating the sum in \eqref{SEM_full_expansion} to contain only 18 terms. \red{The MATLAB code used to generate these results is provided in the supplementary file \texttt{mass\_anchor\_code.mlx}.}

\red{\section{Validation}\label{validation_sec}}
\red{As this paper deals with one-dimensional wave scattering, it is straightforward to validate the results by comparing them to the standard solution. This method uses an ansatz of the form
\begin{equation}\label{standard_solution}
    u(x,t)=\begin{cases}
        F_-\left(t-\frac{x}{c}\right)+H_-\left(t+\frac{x}{c}\right)&x<0\\
        F_+\left(t+\frac{x}{c}\right)+H_+\left(t-\frac{x}{c}\right)&x>0,
    \end{cases}
\end{equation}
i.e., a superposition waves which are incoming ($F_\pm$) and outgoing ($H_\pm$) to the scatterer. After evaluating \eqref{standard_solution} at $x=0$ and applying the kinematic and dynamic conditions of the scatterer (\eqref{kinematic} and \eqref{dynamic}, respectively, for the problem described in \textsection\ref{mass-spring_sec}), the problem reduces to an ordinary differential equation, which we solve numerically. The reader is directed to chapter 1 of \cite{martin2021time} for further details on this standard method.}

\begin{figure}
    \centering
    \includegraphics[width=\textwidth]{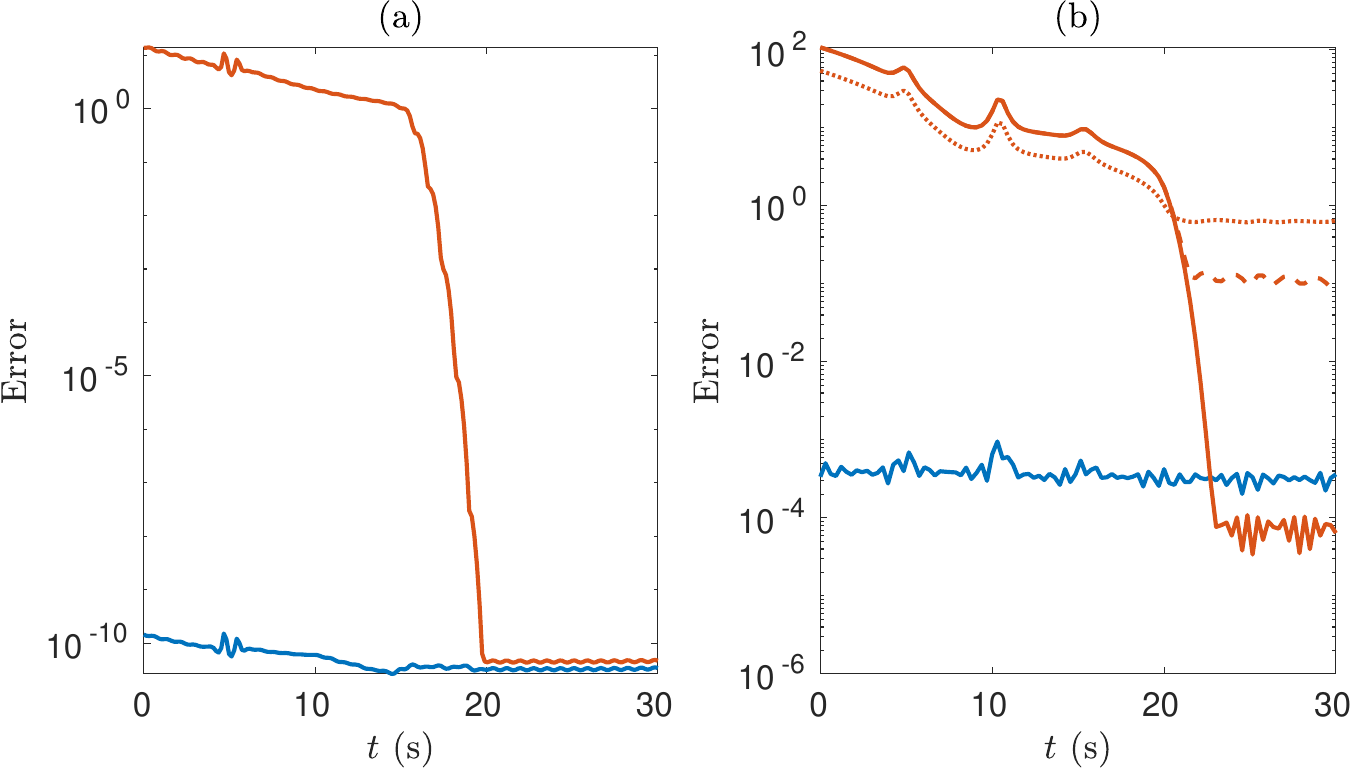}
    \caption{\red{Plots showing the relative error of the GEM (solid blue lines) and SEM (solid red lines) with respect to the standard solution, as defined in the text. Panel (a) shows the error for the problem solved in figure \ref{fig:fig2}, whereas panel (b) shows the error for the problem solved in figure \ref{fig4}, without changing the numerical parameters. Panel (b) also shows the error of the SEM approximation if two (dotted red line) or eight (dashed red line) poles are taken in the SEM expansion.}}
    \label{fig:error_figure}
\end{figure}

\red{To compare the time-dependent error of the GEM relative to the standard solution, we define the error of the GEM solution $u_{\mathrm{GEM}}(x,t)$ as
\begin{align}
    \mathrm{Error}[u_{\mathrm{GEM}}](t) \coloneqq \sqrt{\frac{\sum_{j=1}^{N_x}(u_{\mathrm{GEM}}(x_j,t)-u_{\mathrm{std}}(x_j,t))^2}{\sum_{l=1}^{N_x}u_{\mathrm{std}}(x_l,t)^2}},
\end{align}
i.e., we use quadrature to approximate the relative error between $u_{\mathrm{GEM}}$ and the standard solution $u_{\mathrm{std}}$ over the interval $[x_{\mathrm{min}},x_{\mathrm{max}}]$. A similar quantity to $\mathrm{Error}[u_{\mathrm{GEM}}](t)$ is defined for the SEM. Figure \ref{fig:error_figure} plots these error curves as functions of time for the numerical results presented in figures \ref{fig:fig2} and \ref{fig4}. Observe that the error of the GEM is relatively constant in time, whereas the error of the SEM decreases exponentially after an initial transient. Moreover, panel (b) shows that the SEM becomes more accurate as more poles are taken in the expansion. We remark that the accuracy of the results depends on the initial conditions, as less regular initial conditions would need to be approximated by more complex resonances in the SEM solution, or a larger or more refined frequency discretisation in the GEM solution.}

\section{Conclusion}\label{conclusion_sec}
This paper presented the GEM and SEM in the context of canonical one-dimensional wave scattering. In \textsection\ref{dAlembert_sec}, we introduced the GEM in the context one-dimensional wave propagation on an infinite string and showed that it gives rise to d'Alembert's formula. In \textsection\ref{mass-spring_sec}, we applied the GEM to solve the time domain scattering problem of a mass-spring system attached to a string. After solving the frequency domain problem, we showed that the discrete GEM (which generalises the discrete Fourier transform) can be turned into matrix multiplication after applying a quadrature rule. In \textsection\ref{SEM_sec}, we derived the SEM formula using the Fourier transform, the residue theorem and Steinberg's formula for the residues \cite{steinberg1968meromorphic}. The numerical results shown indicated that the SEM agrees with the GEM for sufficiently large time. Moreover, they indicate that the SEM is accurate at all times if the initial condition is restricted to the resonator---a fact which is due to the underlying Lax-Phillips structure of the problem \cite{pavlov1999irreversibility}. Identical observations were made in the numerical results presented in \textsection\ref{Anchor_point_sec}, where we studied scattering by a mass positioned next to an anchor point. In contrast to the mass-spring system, the complex resonant frequencies of the mass-anchor point problem must be found numerically. Despite only including a small subset of the resonances in the SEM expansion, the agreement with the GEM was not visibly affected.

In its totality, this paper has illustrated two powerful techniques for solving time-domain wave scattering problems. By demonstrating these methods in a canonical wave-scattering context, this paper has made these methods more easily accessible in comparison to existing resources. The GEM and SEM generalise to a wide variety of wave-scattering problems. The application of these methods could lead to new insights into problems that have only been studied in the frequency domain.

\appendix
\section{Orthogonality of frequency domain solutions in the presence of a scatterer}\label{appendix_orthogonality}
In this appendix, we derive equation \eqref{same_normalisation}, namely the result that
\begin{equation*}
    \left\langle\begin{bmatrix}
        \breve{u}_\pm\\ \upi v_\pm
    \end{bmatrix}(\cdot,\omega_1),\begin{bmatrix}
        \breve{u}_\pm\\ \upi v_\pm
    \end{bmatrix}(\cdot,\omega_2)\right\rangle=4\pi\mu c\omega_1\omega_2\delta(\omega_1-\omega_2)
\end{equation*}
in the case of scattering by the mass-spring system. After substituting the frequency domain solutions \eqref{left_incident_FD} and \eqref{right_incident_FD} into \eqref{inner_product_mass_spring}, some rearranging gives
\begin{align}\label{orthogonality_integral_appendix_2}
    \left\langle\begin{bmatrix}
        \breve{u}_\pm\\ \upi v_\pm
    \end{bmatrix}(\cdot,\omega_1),\begin{bmatrix}
        \breve{u}_\pm\\ \upi v_\pm
    \end{bmatrix}(\cdot,\omega_2)\right\rangle&=2\mu\omega_1\omega_2\bigg[\int_{-\infty}^0 e^{\upi(k_1-k_2)x}+r(\omega_1)r(\omega_2)^*e^{\upi(k_2-k_1)x}\upd x\nonumber\\
    &\qquad+\int_0^\infty t(\omega_1)t(\omega_2)^*e^{\upi(k_1-k_2)x}\upd x\bigg]\nonumber\\
    &\qquad+(M\omega_1\omega_2+k_s)t(\omega_1)t(\omega_2)^*.
\end{align}
To compute the integrals, we use the formula for the Fourier transform of the Heaviside step function to derive the following:
\begin{subequations}
    \begin{align}
    \int_0^\infty e^{\upi\omega t}\upd t&=\pi\delta(\omega)+\frac{\upi}{\omega}\\
    \int_{-\infty}^0 e^{\upi\omega t}\upd t&=\pi\delta(\omega)-\frac{\upi}{\omega}.
\end{align}
\end{subequations}
By applying these and the conservation of energy identity ${|r(\omega)|^2+|t(\omega)|^2=1}$, \eqref{orthogonality_integral_appendix_2} becomes
\begin{align}\label{orthogonality_integral_appendix_3}
    \left\langle\begin{bmatrix}
        \breve{u}_\pm\\ \upi v_\pm
    \end{bmatrix}(\cdot,\omega_1),\begin{bmatrix}
        \breve{u}_\pm\\ \upi v_\pm
    \end{bmatrix}(\cdot,\omega_2)\right\rangle&=4\pi\mu c\omega_1^2\delta(\omega_1-\omega_2)\nonumber\\
    &\quad+\frac{2\mu\omega_1\omega_2(1-r(\omega_1)r(\omega_2)^*-t(\omega_1)t(\omega_2)^*)}{\upi(k_1-k_2)}\nonumber\\
    &\quad+(M\omega_1\omega_2+k_s)t(\omega_1)t(\omega_2)^*.
\end{align}
After some lengthy algebra, substitution of the expressions for the reflection and transmission coefficients \eqref{r_and_t_coeff} into \eqref{orthogonality_integral_appendix_3} reveals that the second and third terms of \eqref{orthogonality_integral_appendix_3} cancel, which leaves us with the desired result.

\bibliographystyle{elsarticle-num} 
\bibliography{bibfile}

\begin{thebibliography}{10}
\expandafter\ifx\csname url\endcsname\relax
  \def\url#1{\texttt{#1}}\fi
\expandafter\ifx\csname urlprefix\endcsname\relax\def\urlprefix{URL }\fi
\expandafter\ifx\csname href\endcsname\relax
  \def\href#1#2{#2} \def\path#1{#1}\fi

\bibitem{martin2006}
P.~Martin, Multiple scattering: interaction of time-harmonic waves with N obstacles, no. 107 in Encyclopedia of Mathematics and its Applications, Cambridge University Press, 2006.

\bibitem{povzner1953expansion}
A.~Y. Povzner, On the expansion of arbitrary functions in characteristic functions of the operator -$\delta u+cu$, Matematicheskii Sbornik 74~(1) (1953) 109--156, (in Russian).

\bibitem{ikebe1960eigenfunction}
T.~Ikebe, Eigenfunction expansions associated with the schroedinger operators and their applications to scattering theory, Archive for Rational Mechanics and Analysis 5 (1960) 1--34.

\bibitem{hazard2007generalized}
C.~Hazard, F.~Loret, Generalized eigenfunction expansions for conservative scattering problems with an application to water waves, Proceedings of the Royal Society of Edinburgh Section A: Mathematics 137~(5) (2007) 995--1035.
\newblock \href {https://doi.org/10.1017/S0308210506000138} {\path{doi:10.1017/S0308210506000138}}.

\bibitem{hazard2008spectral}
C.~Hazard, M.~H. Meylan, Spectral theory for an elastic thin plate floating on water of finite depth, SIAM Journal on Applied Mathematics 68~(3) (2008) 629--647.
\newblock \href {https://doi.org/10.1137/060665208} {\path{doi:10.1137/060665208}}.

\bibitem{meylan2009time}
M.~H. Meylan, R.~E. Taylor, Time-dependent water-wave scattering by arrays of cylinders and the approximation of near trapping, Journal of Fluid Mechanics 631 (2009) 103--125.
\newblock \href {https://doi.org/10.1017/S0022112009007204} {\path{doi:10.1017/S0022112009007204}}.

\bibitem{meylan_2009}
M.~H. Meylan, Time-dependent linear water-wave scattering in two dimensions by a generalized eigenfunction expansion, Journal of Fluid Mechanics 632 (2009) 447–455.
\newblock \href {https://doi.org/10.1017/S002211200900723X} {\path{doi:10.1017/S002211200900723X}}.

\bibitem{peter2010general}
M.~A. Peter, M.~H. Meylan, A general spectral approach to the time-domain evolution of linear water waves impacting on a vertical elastic plate, SIAM Journal on Applied Mathematics 70~(7) (2010) 2308--2328.
\newblock \href {https://doi.org/10.1137/090756557} {\path{doi:10.1137/090756557}}.

\bibitem{meylan_fitzgerald_2014}
M.~H. Meylan, C.~J. Fitzgerald, The singularity expansion method and near-trapping of linear water waves, Journal of Fluid Mechanics 755 (2014) 230–250.
\newblock \href {https://doi.org/10.1017/jfm.2014.411} {\path{doi:10.1017/jfm.2014.411}}.

\bibitem{Meylan2023MWSW03}
M.~H. Meylan, \href{https://www.newton.ac.uk/seminar/38986/}{Time-domain calculations for multiple wave scattering}, presentation given at Isaac Newton Institute Multiple Wave Scattering Workshop 3: Computational methods for multiple scattering (2023).
\newline\urlprefix\url{https://www.newton.ac.uk/seminar/38986/}

\bibitem{martin2021time}
P.~A. Martin, Time-domain scattering, Vol. 180, Cambridge University Press, 2021.

\bibitem{baum1971singularity}
C.~E. Baum, \href{https://apps.dtic.mil/sti/citations/ADA066905}{On the singularity expansion method for the solution of electromagnetic interaction problems}, Tech. rep., Air Force Weapons Lab Kirtland (1971).
\newline\urlprefix\url{https://apps.dtic.mil/sti/citations/ADA066905}

\bibitem{baum2005singularity}
C.~E. Baum, The singularity expansion method, in: L.~B. Felsen (Ed.), Transient electromagnetic fields, Springer, 2005, pp. 129--179.

\bibitem{Baum_1997}
C.~E. Baum, Discrimination of buried targets via the singularity expansion, Inverse Problems 13~(3) (1997) 557.
\newblock \href {https://doi.org/10.1088/0266-5611/13/3/003} {\path{doi:10.1088/0266-5611/13/3/003}}.

\bibitem{pagneux2013trapped}
V.~Pagneux, Trapped modes and edge resonances in acoustics and elasticity, in: R.~Craster, J.~Kaplunov (Eds.), Dynamic Localization Phenomena in Elasticity, Acoustics and Electromagnetism, Springer, 2013, pp. 181--223.

\bibitem{Kristensen2020}
P.~T. Kristensen, K.~Herrmann, F.~Intravaia, K.~Busch, Modeling electromagnetic resonators using quasinormal modes, Adv. Opt. Photon. 12~(3) (2020) 612--708.
\newblock \href {https://doi.org/10.1364/AOP.377940} {\path{doi:10.1364/AOP.377940}}.

\bibitem{meylan_2002}
M.~H. Meylan, Spectral solution of time-dependent shallow water hydroelasticity, Journal of Fluid Mechanics 454 (2002) 387–402.
\newblock \href {https://doi.org/10.1017/S0022112001007273} {\path{doi:10.1017/S0022112001007273}}.

\bibitem{baldassari2021modal}
L.~Baldassari, P.~Millien, A.~L. Vanel, Modal approximation for plasmonic resonators in the time domain: the scalar case, Partial Differential Equations and Applications 2~(4) (2021) 46.
\newblock \href {https://doi.org/10.1007/s42985-021-00098-4} {\path{doi:10.1007/s42985-021-00098-4}}.

\bibitem{billingham2000wave}
J.~Billingham, A.~C. King, Wave motion, no.~24 in Cambridge Texts in Applied Mathematics, Cambridge University Press, 2000.

\bibitem{MARTIN20142_N_masses}
P.~Martin, N masses on an infinite string and related one-dimensional scattering problems, Wave Motion 51~(2) (2014) 296--307.
\newblock \href {https://doi.org/10.1016/j.wavemoti.2013.08.005} {\path{doi:10.1016/j.wavemoti.2013.08.005}}.

\bibitem{MARTIN2015_semi-infinite}
P.~Martin, I.~D. Abrahams, W.~J. Parnell, One-dimensional reflection by a semi-infinite periodic row of scatterers, Wave Motion 58 (2015) 1--12.
\newblock \href {https://doi.org/10.1016/j.wavemoti.2015.06.005} {\path{doi:10.1016/j.wavemoti.2015.06.005}}.

\bibitem{porter2018waves}
R.~Porter, \href{https://people.maths.bris.ac.uk/~marp/abstracts/string.pdf}{Waves on a long string with side branches}, Tech. rep., University of Bristol (2018).
\newline\urlprefix\url{https://people.maths.bris.ac.uk/~marp/abstracts/string.pdf}

\bibitem{davies2023problem}
B.~Davies, L.~Fehertoi-Nagy, H.~J. Putley, On the problem of comparing graded metamaterials, Proceedings of the Royal Society A: Mathematical, Physical and Engineering Sciences 479~(2277) (2023) 20230537.
\newblock \href {https://doi.org/10.1098/rspa.2023.0537} {\path{doi:10.1098/rspa.2023.0537}}.

\bibitem{bennetts2021complex}
L.~G. Bennetts, M.~H. Meylan, Complex resonant ice shelf vibrations, SIAM Journal on Applied Mathematics 81~(4) (2021) 1483--1502.
\newblock \href {https://doi.org/10.1137/20M1385172} {\path{doi:10.1137/20M1385172}}.

\bibitem{lamb1900peculiarity}
H.~Lamb, On a peculiarity of the wave-system due to the free vibrations of a nucleus in an extended medium, Proceedings of the London Mathematical Society 1~(1) (1900) 208--213.
\newblock \href {https://doi.org/10.1112/plms/s1-32.1.208} {\path{doi:10.1112/plms/s1-32.1.208}}.

\bibitem{wilcox1975scattering}
C.~H. Wilcox, Scattering theory for the d'Alembert equation in exterior domains, Vol. 442, Springer-Verlag, 1975.

\bibitem{hazard2002}
C.~Hazard, M.~Lenoir, Surface water waves, in: R.~Pike, P.~Sabatier (Eds.), Scattering, Vol.~1, Academic Press, 2002, pp. 618--636.

\bibitem{steinberg1968meromorphic}
S.~Steinberg, Meromorphic families of compact operators, Archive for Rational Mechanics and Analysis 31 (1968) 372--379.
\newblock \href {https://doi.org/10.1007/BF00251419} {\path{doi:10.1007/BF00251419}}.

\bibitem{meylan2012complex}
M.~H. Meylan, M.~Tomic, Complex resonances and the approximation of wave forcing for floating elastic bodies, Applied Ocean Research 36 (2012) 51--59.
\newblock \href {https://doi.org/10.1016/j.apor.2012.02.003} {\path{doi:10.1016/j.apor.2012.02.003}}.

\bibitem{pavlov1999irreversibility}
B.~Pavlov, Irreversibility, {Lax-Phillips} approach to resonance scattering and spectral analysis of non-self-adjoint operators in {Hilbert} space, International Journal of Theoretical Physics 38~(1) (1999) 21--45.
\newblock \href {https://doi.org/10.1023/A:1026624905808} {\path{doi:10.1023/A:1026624905808}}.

\end{thebibliography}

\end{document}